\documentclass[journal]{IEEEtran}
\usepackage{amsmath,amsfonts}
\usepackage{algorithmic}
\usepackage{algorithm}
\usepackage{array}
\usepackage[caption=false,font=normalsize,labelfont=sf,textfont=sf]{subfig}
\usepackage{textcomp}
\usepackage{stfloats}
\usepackage{url}
\usepackage{verbatim}
\usepackage{graphicx}
\usepackage{cite}
\usepackage{bm}
\usepackage{amssymb}
\usepackage{multirow}     
\usepackage{diagbox}    
\usepackage{array}      
\usepackage{colortbl}
\usepackage{longtable}   
\usepackage{xcolor}
\usepackage{xpatch}
\definecolor{ChangeColor}{rgb}{1,0,0}
\def\BibTeX{{\rm B\kern-.05em{\sc i\kern-.025em b}\kern-.08em
   T\kern-.1667em\lower.7ex\hbox{E}\kern-.125emX}}
\begin{document}

\title{Full-Dimensional Beamforming for Multi-User MIMO-OFDM ISAC for Low-Altitude UAV with Zero Sensing Resource Allocation}

\author{Zhiwen~Zhou,
        Yong~Zeng,~\IEEEmembership{Fellow,~IEEE},
        Chunguo Li,~\IEEEmembership{Senior Member,~IEEE},
        Fei Yang,
        Yan Chen,~\IEEEmembership{Senior Member,~IEEE},
        and 
        Jingon Joung,~\IEEEmembership{Senior Member,~IEEE}
        \vspace{-2em}
\thanks{This work was supported by the Natural Science Foundation for Distinguished Young Scholars of Jiangsu Province with grant number BK20240070. 

Z. Zhou and Y. Zeng are with the National Mobile Communications Research Laboratory, Southeast University, Nanjing 210096, China. Y. Zeng is also with the Purple Mountain Laboratories, Nanjing 211111, China (e-mail: {zhiwen\_zhou, yong\_zeng}@seu.edu.cn).

C. Li is with the School of Information Science and Engineering, Southeast University, Nanjing 210096, China (e-mail: chunguoli@seu.edu.cn).

F. Yang and Y. Chen are with the Wireless Technology Lab., Huawei Technologies Co., Ltd, Shanghai 201206, China. (e-mail: {yangfei8, bigbird.chenyan}@huawei.com).

J. Joung is with the School of Electrical and Electronics Engineering, Chung-Ang University, Seoul 06974, South Korea (e-mail: jgjoung@cau.ac.kr).
}
}

\maketitle

\begin{abstract}
Low-altitude unmanned aerial vehicles (UAVs) are expected to play an important role for low-altitude economy with a wide range of applications like precise agriculture, aerial delivery and surveillance. Integrated sensing and communication (ISAC) is a key technology to enable the large-scale deployment and routine usage of UAVs by providing both communication and sensing services efficiently. For UAV ISAC systems, as UAV often acts as both a communication user equipment (UE) and a sensing target, traditional ISAC systems that usually allocate dedicated TF resources for sensing are inefficient due to the severe degradation of communication spectral efficiency. To address this issue, in this paper, we propose a novel multiple-input multiple-output (MIMO) orthogonal frequency division multiplexing (OFDM)-based ISAC framework for UAVs that eliminates the need for dedicated sensing TF resources, achieving zero TF sensing overhead. By designing the transmit beamforming to meet the requirements for both communication and sensing tasks, our proposed approach enables the communication TF resources to be fully reused for sensing, thereby enhancing both the communication sum rate and the sensing performance in terms of resolution, unambiguous range, and accuracy. Additionally, we introduce a low-complexity target searching beamforming algorithm and a two-stage super-resolution sensing algorithm, which ensure efficient implementation. Simulation results demonstrate that the proposed MIMO-OFDM-ISAC framework not only improves the communication sum rate but also outperforms traditional ISAC systems in sensing performance, making it a promising solution for future ISAC systems to support low-altitude UAVs.
\end{abstract}


\IEEEpeerreviewmaketitle
\section{Introduction}
Low-altitude unmanned aerial vehicles (UAVs) are expected to revolutionize various fields, particularly for the emerging low-altitude economy with a wide range of applications including precise agriculture, aerial delivery and surveillance. As the applications of UAVs continue to expand, there is a growing need for mobile networks to support their unique requirements. In particular, UAVs require enhanced communication capabilities to cater to their higher altitudes and dynamic movements while delivering high capacity, low-latency and ultra-reliable wireless connectivity for not only payload communications, but also control-and-nonpayload communications (CNPC) \cite{8918497}. Furthermore, powerful sensing capabilities are crucial not only for aerial traffic management to timely detect and track non-cooperative or even malicious UAVs, but also for cooperative UAVs to maintain precise positional and environmental awareness \cite{song2024overview}.

Integrated sensing and communication (ISAC) emerges as a pivotal technology to support the large-scale deployment and routine usage of UAVs. By seamlessly combining communication and sensing functionalities, ISAC not only adds sensing capabilities to mobile communication networks, but also facilitates enhanced communication, including 3D beam tracking and dynamic channel estimation\cite{dai2025tutorial}. As a result, ISAC has been identified as one of the six typical usage scenarios for the sixth generation (6G) mobile communication networks \cite{sector2023framework}

One of the key challenges in ISAC systems is the efficient allocation of radio resources between communication and sensing tasks\cite{xiao2024achieving,9945983}. Traditional ISAC systems often utilize dedicated signals for sensing, i.e., a portion of the time-frequency (TF) resources are exclusively allocated for sensing. For example, current fifth-generation-advanced (5G-A) ISAC experimental systems deployed by China Mobile Communications Group (CMCC) typically allocate 10\% to 20\% TF resources for sensing. With orthogonal resource allocation between sensing and communication, it reduces the available resources for communication and limits the overall system performance\cite{9724187}. On the other hand, with non-orthogonal resource allocation, interference arises and it requires extra procedure to suppress these interference\cite{10770127,10086626,10158711}.
Orthogonal frequency division multiplexing (OFDM) is a widely adopted modulation scheme in modern wireless communication systems due to its robustness against multipath fading, high spectral efficiency, and high flexibility for resource allocation. It is also considered highly favorable for ISAC due to its convenience for delay-Doppler sensing estimation as well as low ranging sidelobe properties \cite{dai2025tutorial,liu2024ofdm}. The CP duration required for inter-symbol interference (ISI)-free monostatic sensing might increase, however, this can be mitigated with the sliding window approach proposed in \cite{xu2025does}. Many existing works on OFDM-based ISAC systems focus on orthogonal TF resource allocation or relying on pilot signals for sensing. For instance, \cite{10693761} investigated the design of TF resource patterns for reference signals in OFDM systems; \cite{10570265,10626507} aimed to reduce the sensing overhead by leveraging coprime-based slot and subcarrier allocation; \cite{xiao2024achieving} proposed a two-step resource allocation scheme, dynamically adjusting the resource allocation in the second step to achieve superior sensing performance with limited radio resources. \cite{zhang2024prototyping} developed an OFDM ISAC prototyping system based on dedicated sensing time-slots for environment sensing. The use of dedicated signals or pilots for sensing is primarily motivated by its simplicity and ease of implementation. Sacrificially, for multi-user multiple-input multiple-output (MU-MIMO) communication systems, since user data signals are typically beamformed towards the user's direction, only targets located near the user can be sensed if these signals are directly reused for sensing. Therefore, by employing dedicated signals and pilots, no additional design modifications are required for the existing MU-MIMO operations. 

However, in the context of ISAC for low-altitude UAVs, UAVs may not only act as communication user equipments (UEs) but also serve as sensing targets. In this case, as the communication UEs and sensing targets are not sufficiently isolated in the spatial domain, orthogonal TF resource allocation is typically used to avoid mutual interference between communication and sensing signals\cite{10770127}, which, however, would lead to severe under-utilization of TF resources and redundant power consumption, ultimately degrading spectral efficiency and sensing performance. The dual role of low-altitude UAVs suggests that the wireless resources used for communication with one UAV can also be effectively reused for sensing itself and its surrounding environment. Motivated by this observation, we  aim to develop more efficient designs of MIMO-OFDM ISAC systems for UAVs. 

The utilization of communication TF resources for sensing has been investigated in early studies on OFDM ISAC. However, these studies typically consider scenarios where the sensing target and communication user are either co-located or in close proximity, or assume omnidirectional transmission where the signals are broadcasted uniformly in all directions \cite{4977002,2014ofdmradar,5073387,braun2012usrp}.
There has also been research considering reusing communication TF resources for sensing in MIMO-OFDM ISAC systems, exploring beamforming and precoder design. For instance, \cite{10622793} proposed a beamforming design tailored for multi-user MIMO-OFDM ISAC systems. Their approach optimizes beamforming to balance the trade-off between the peak-to-sidelobe ratio (PSLR) of the range-Doppler (RD) map and the sum capacity. In another study, \cite{10001164} presented a joint design of dual-functional transmit signals across multiple subcarriers, enabling multi-user OFDM communications while simultaneously detecting a moving target in the presence of clutter. Their beamforming strategy seeks to maximize the sensing signal-to-interference-plus-noise ratio (SINR) while ensuring a minimum communication SINR. \cite{10255745} investigated the waveform design for MIMO-OFDM ISAC systems from a information theory perspective and proposed a closed-form expressions for efficient realization. \cite{xu2023bandwidth} introduced a novel method that first utilizes shared subcarriers for coarse angle estimation, followed by private subcarriers transmitting orthogonal pilots across different transmit antennas to form a virtual array. This approach achieves more accurate angle estimation at the expense of reduction in communication rate, with the beamforming design for shared subcarriers balancing beam pattern error and sensing signal-to-noise ratio (SNR). Additionally, \cite{10736664} proposed an innovative compressed sensing algorithm to achieve accurate estimation with a significantly reduced number of subcarriers, complemented by a dedicated beamforming strategy to further enhance its performance.

However, the aforementioned works primarily focus on specific aspects, such as beamforming design or sensing algorithms, without providing a comprehensive system design framework for multi-user MIMO OFDM ISAC. In particular, the beamforming optimization and sensing algorithms proposed in these studies face new challenges when directly applied to ISAC for UAVs. This is primarily due to the need to support 3D spatial domain ISAC, which inevitably complicates the beamforming design. Additionally, ISAC for UAVs involves 4D parameter estimation, including azimuth and elevation angles, ranges, and radial velocities. In such cases, the existing joint parameter estimation algorithms will become overly complex and computationally demanding, limiting their practical applicability.

To address the above issues, in this paper, we propose a novel multi-user MIMO OFDM-based ISAC framework for low-altitude UAVs, which eliminates the need for dedicated sensing TF resources, achieving zero sensing overhead. The key idea is to fully reuse the communication signals for sensing purposes by designing the transmit beamforming. This approach allows the system to enhance both the communication throughput and the sensing performance in terms of resolution, unambiguous sensing range, and sensing accuracy. To ensure efficient implementation, we propose a low-complexity searching-stage beamforming algorithm and a two-stage super-resolution sensing algorithm. Unlike traditional ISAC systems, which sacrifice communication resources for sensing, our proposed method leverages the full potential of the available resources, achieving superior performance in both domains.

The main contributions of this paper are as follows:
\begin{itemize}
\item Zero Sensing Overhead: We propose a novel OFDM-ISAC framework tailored for low-altitude UAV applications that eliminates the need for dedicated sensing TF resources, thereby maximizing the communication throughput while improving sensing performance.

\item Beamforming Design: We develop a transmit beamforming strategy that ensures both communication quality and sensing performance. The beamforming vectors are designed to minimize the transmit power while meeting the requirements for communication and sensing tasks. Furthermore, we introduce a closed-form and optimization-free beamforming algorithm for the searching stage, enabling low-complexity implementation.

\item Sensing Algorithms: To address the unique challenges of ISAC for UAVs, we propose a low-complexity two-stage super-resolution sensing algorithm. In the first stage, 2D angle estimation is performed using the multiple signal classification (MUSIC) algorithm to achieve high angular resolution. In the second stage, 2D delay-Doppler estimation is conducted for each estimated spatial angle, avoiding the computational complexity of 4D joint estimation. To further enhance the MUSIC algorithm's performance under low SNR conditions typical in ISAC for UAVs, we propose a novel denoising algorithm prior to angle estimation.

\item Performance Analysis: We conduct a comprehensive comparison of the proposed ISAC system for UAVs against traditional ISAC systems with dedicated sensing TF resources. The results demonstrate that our proposed approach not only improves communication performance but also outperforms traditional ISAC systems in key sensing metrics. This is attributed to the efficient utilization of all TF resources for both sensing and communication.
\end{itemize}

The remainder of this paper is organized as follows. 
Section \ref{sysmodel} introduces the system model of a multi-user MIMO-OFDM ISAC system for low-altitude UAVs. Section \ref{sec_resource} compares the resource allocation schemes with dedicated and zero sensing resources.
The beamforming designs for target searching and tracking with dedicated and zero sensing resource allocation are discussed in Section \ref{Beamforming Design} while 
Section \ref{Sensing Algo} presents the corresponding sensing algorithms. Simulation results are discussed in Section \ref{sim_results}. Finally, we conclude this paper in Section \ref{conclusion}.

\textit{Notation}: Italic, bold-faced, lower- and upper- case characters denote scalars, vectors, and matrices, respectively. The transpose, Hermitian transpose, and complex conjugate operation are given by $(\cdot)^T$, $(\cdot)^H$, and $(\cdot)^*$, respectively. $\mathbb{C}^{M\times N}$ and $\mathbb{R}^{M\times N}$ signify the spaces of $M\times N$ complex and real matrices. $j = \sqrt{-1}$ denotes the imaginary unit of complex numbers. The distribution of a circularly symmetric complex Gaussian (CSCG) random variable with mean 0 and variance $\sigma^{2}$ is denoted by $\mathcal{CN}(0,\sigma^{2})$. $\lceil\cdot\rceil$ and $\lfloor\cdot\rfloor$ denote the ceiling and floor operations, respectively. $\otimes$ denotes the Kronecker product operation. When used with a set, $|\cdot|$ denotes its cardinality. $B\backslash A$ denotes the complement of set $A$ with respect to set $B$.

\section{System Model}
\label{sysmodel}
\begin{figure}[!htbp]
\vspace*{-8pt}
\centering
\includegraphics[width=0.392\textwidth]{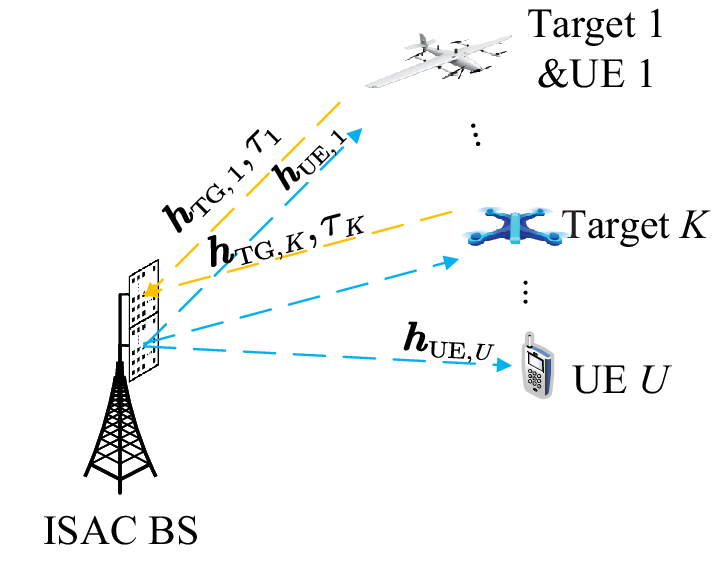}
\vspace*{-6pt}
\caption{A multi-user MIMO-OFDM ISAC system for low-altitude UAVs with $U$ communication users and $K$ sensing targets.}
\label{scene}
\vspace*{-6pt}
\end{figure}
As shown in Fig. \ref{scene}, we consider a downlink multi-user MIMO-OFDM ISAC system with $U$ communication UEs and $K$ sensing targets, where each UE has one antenna and the BS is equipped with two $M\times N$ uniform planar array (UPAs) for downlink ISAC signal transmission and uplink sensing signal reception, respectively, with $M$ and $N$ denoting the number of antenna elements in the horizontal and vertical directions, respectively. The choice of UPAs over traditional uniform linear arrays (ULAs) is motivated by the need to support full-dimensional MIMO ISAC for low-altitude UAVs. It should be noted that the sensing targets can also be communication UEs, as demonstrated in Fig. \ref{scene} where target $1$ and UE $1$ corresponds to the same UAV. The reception (RX) UPA is positioned directly above the transmitter (TX) UPA, with adequate separation to maintain TX-RX isolation, while ensuring that the angle of arrival (AoA) and angle of departure (AoD) for the same sensing target are equal. The channel impulse response between the ISAC BS and the $u$th UE can be written as
\begin{equation}
\abovedisplayshortskip=2pt
\belowdisplayshortskip=2pt
\abovedisplayskip=2pt
\belowdisplayskip=2pt
\begin{aligned}
\boldsymbol{h}_{\mathrm{UE},u}\left( t,\tau \right) =\sum_{l=1}^{L_u}{\alpha_{u,l}\boldsymbol{a}\left( \phi _{u,l},\theta_{u,l} \right) \delta \left( \tau -\tau _{u,l} \right) e^{j2\pi f_{D,u,l}t}},
\end{aligned}
\label{ComChannel}
\end{equation}
where $u\in\{1,\ldots, U\}$ is the UE index, $L_u$ is the number of propagation paths between the BS and the $u$th UE, with $\alpha_{u,l}$, $\boldsymbol{a}\left( \phi _{u,l},\theta_{u,l} \right)$, $\tau_{u,l}$ and $f_{D,u,l}$ denoting the complex scattering coefficient, steering vector, propagation delay and Doppler shift of the $l$th path for UE $u$, respectively.
For the $M\times N$-element UPA used by the BS, the steering vector can be expressed as
\begin{equation}
\abovedisplayshortskip=2pt
\belowdisplayshortskip=2pt
\abovedisplayskip=2pt
\belowdisplayskip=2pt
\begin{aligned}
\boldsymbol{a}\left( \phi ,\theta \right) =\boldsymbol{a}_x\left( \phi ,\theta \right) \otimes \boldsymbol{a}_z\left( \theta \right) \in \mathbb{C} ^{MN\times 1}.
\end{aligned}
\label{SteeringVec2D}
\end{equation}
$\boldsymbol{a}_x\left( \phi ,\theta \right)$ and $\boldsymbol{a}_z\left(\theta \right)$ are expressed as
\begin{equation}
\abovedisplayshortskip=2pt
\belowdisplayshortskip=2pt
\abovedisplayskip=2pt
\belowdisplayskip=2pt
\begin{aligned}
&\boldsymbol{a}_x\left( \phi ,\theta \right) =\left[ e^{j2\pi \frac{md\cos \phi \sin \theta}{\lambda}} \right] _{m=0}^{M-1}\in \mathbb{C} ^{M\times 1},
\\
&\boldsymbol{a}_z\left( \theta \right) =\left[ e^{j2\pi \frac{nd\cos \theta}{\lambda}} \right] _{n=0}^{N-1}\in \mathbb{C} ^{N\times 1},
\end{aligned}
\label{SteeringVec}
\end{equation}
where $d$ is the element spacing, $\lambda$ is the wavelength, $\phi$ is the azimuth angle and $\theta$ is the zenith angle. The monostatic sensing MIMO channel between the TX-UPA and RX-UPA of the BS via the $k$th sensing target can be written as
\begin{equation}
\abovedisplayshortskip=2pt
\belowdisplayshortskip=2pt
\abovedisplayskip=2pt
\belowdisplayskip=2pt
\begin{aligned}
\boldsymbol{H}_{\mathrm{TG},k}\left( t,\tau \right) =\alpha _k\boldsymbol{a}_k \boldsymbol{a}_k^T \delta \left( \tau -\tau _k \right) e^{j2\pi f_{D,k}t},
\end{aligned}
\label{SensingChannel}
\end{equation}
in which $\boldsymbol{a}_k\triangleq \boldsymbol{a}\left( \phi _k,\theta _k \right)$, $k\in\{1,\ldots, K\}$ is the target index, $\alpha_k$, $\boldsymbol{a }\left( \phi _k, \theta_k \right)$, $\tau_k = 2R_k/c$ and $f_{D,k}=2v_k/\lambda $ are the scattering coefficient, steering vector, propagation delay and Doppler shift of the $k$th target respectively, with $\phi_k$, $\theta_k$, $R_k$ and $v_k$ denoting the azimuth and zenith angle, range and radial velocity of target $k$, respectively. Since the energy of higher-order reflections is relatively weak, only the first-order reflections are considered.

For communication, we consider the widely-adopted OFDMA scheme for multiple access among the $U$ UEs. A common practice for converting the traditional OFDMA communication system into an ISAC system is to allocate dedicated TF resources for sensing, i.e., the signal $\boldsymbol{s}(t)\in \mathbb{C}^{MN\times 1}$ transmitted by the BS is given by
\begin{equation}
\abovedisplayshortskip=2pt
\belowdisplayshortskip=2pt
\abovedisplayskip=2pt
\belowdisplayskip=2pt
\begin{aligned}
\boldsymbol{s}\left( t \right) =\boldsymbol{s}_c\left( t \right) +\boldsymbol{s}_s\left( t \right),
\end{aligned}
\label{Sig_Tx}
\end{equation}
where $\boldsymbol{s}_c\left( t \right)$ is the communication signal given by
\begin{equation}
\abovedisplayshortskip=2pt
\belowdisplayshortskip=2pt
\abovedisplayskip=2pt
\belowdisplayskip=2pt
\begin{aligned}
&\boldsymbol{s}_c\left( t \right) =\sum_{u=0}^{U-1}{\sum_{\left( {p},{q} \right) \in \varGamma _u}{\boldsymbol{f}_{u,{q}}b_{{p},{q}}}}
\\
&\cdot e^{j2\pi {p}\Delta f\left( t\!-{q}T_O\!-\!T_{\mathrm{CP}} \right)}\mathrm{rect}\!\left( \!\frac{t\!-{q}T_O}{T_O}\! \right),
\end{aligned}
\label{Sig_sensing}
\end{equation}
and $\boldsymbol{s}_s\left( t \right)$ is the dedicated sensing signal given by
\begin{equation}
\abovedisplayshortskip=2pt
\belowdisplayshortskip=2pt
\abovedisplayskip=2pt
\belowdisplayskip=2pt
\begin{aligned}
&\boldsymbol{s}_s\left( t \right) =\sum_{\left( {p},{q} \right) \in \varGamma _s}{\boldsymbol{f}_{s,{q}}b_{{p},{q}}}
\\
&\cdot e^{j2\pi {p}\Delta f\left( t\!-{q}T_O\!-\!T_{\mathrm{CP}} \right)}\mathrm{rect}\!\left( \!\frac{t\!-{q}T_O}{T_O}\! \right).
\end{aligned}
\label{Sig_sensing1}
\end{equation}
In \eqref{Sig_sensing} and \eqref{Sig_sensing1}, $\Delta f$ is the subcarrier spacing, $T_{\mathrm{CP}}$ is the duration of the cyclic prefix (CP) and $T_O$ is the duration of the OFDM symbol including CP. $\{b_{{p},{q}}\}_{{p}=0,{q}=0}^{{p}={P}-1,{q}={Q}-1}$ contains ${P}{Q}$ independent and identically distributed (i.i.d) modulation symbols, where element $b_{{p},{q}}$ represents the symbol on the ${p}$th subcarrier and ${q}$th OFDM symbol with $\mathbb{E} \left\{ \left|b_{{p},{q}} \right|^2 \right\} =1/{P}$. $\varGamma _u$ denotes the set of resource elements allocated to user $u$ with $\varGamma _i\cap \varGamma _j=\varnothing , \forall i\ne j$, and $\varGamma _s$ denotes the set of dedicated resource elements for sensing. Furthermore, $\boldsymbol{f}_{u,{q}}$ and $\boldsymbol{f}_{s,{q}}$ denote the corresponding beamforming vectors for all the subcarriers at the ${q}$th OFDM symbol. Note that $\varGamma _s\cup \bigcup_{u=0}^{U-1}{\varGamma _u}=\varGamma $, where $\varGamma$ denotes the set of all the available resource elements. 

Based on \eqref{ComChannel}, the signal received by the $u$th UE can be expressed as,
\begin{equation}
\abovedisplayshortskip=2pt
\belowdisplayshortskip=2pt
\abovedisplayskip=2pt
\belowdisplayskip=2pt
\begin{aligned}
&y_u\left( t \right) =\int_{-\infty}^{\infty}{\boldsymbol{h}_{\mathrm{UE},u}^{T}\left( t,\tau \right) \boldsymbol{s}\left( t-\tau \right)d\tau}+z_u\left( t \right) 
\\
&=\sum_{l=1}^{L_u}{\alpha _{u,l}\boldsymbol{a}_{u,l}^T \boldsymbol{s}\left( t-\tau _{u,l} \right) e^{j2\pi f_{D,u,l}t}}+z_u\left( t \right),
\end{aligned}
\label{Sig_r_u}
\end{equation}
where $\boldsymbol{a}_{u,l}\triangleq \boldsymbol{a}\left( \phi _{u,l},\theta _{u,l} \right) $, $z_u(t) \sim \mathcal{CN}(0, \sigma^2)$ represents the additive Gaussian white noise, with $\sigma^2 = B N_0  $ denoting the noise power. Here, $B = P \Delta f$ is the signal bandwidth, and $N_0$ is the double-sided noise power spectral density. The $u$th UE synchronizes to its strongest path and demodulates the OFDM signal. Without loss of generality, we assume that the first path $l=1$ is the strongest among the $L_u$ paths, and further denote the delays and Doppler shifts after synchronization by $\tilde{\tau}_{u,l}=\tau_{u,l}-{\tau}_{u,1}$ and $\tilde{f}_{D,u,l}=f_{D,u,l}-f_{D,u,1}$. Assuming that the delay spread is smaller than $T_{\mathrm{CP}}$ and the Doppler spread is small enough so that the ICI is negligible, the received resource grid $\boldsymbol{Y}_u\in\mathbb{C} ^{{P}\times {Q}}$ can be written as
\begin{equation}
\abovedisplayshortskip=2pt
\belowdisplayshortskip=2pt
\abovedisplayskip=2pt
\belowdisplayskip=2pt
\begin{aligned}
\left[ \boldsymbol{Y}_u \right] _{p,q}&=\sum_{l=1}^{L_u}{\alpha _{u,l}\boldsymbol{a}_{u,l}^T e^{j2\pi \left( \tilde{f}_{D,u,l}qT_O-p\Delta f\tilde{\tau}_{u,l} \right)}}\boldsymbol{f}_{u,q}b_{p,q}
\\
&+\left[ \boldsymbol{Z}_u \right] _{p,q},\left( p,q \right) \in \varGamma _u,
\end{aligned}
\label{Res_r_u}
\end{equation}
where $[ \boldsymbol{Y}_u ] _{{p},{q}}$ denotes the element on the ${p}$th row and ${q}$th column of matrix $\boldsymbol{Y}_u$, and $[ \boldsymbol{Z}_u ] _{p,q} \sim \mathcal{CN} ( 0,\tilde{\sigma} ^2 )$ represents the additive Gaussian noise at the corresponding resource grid element, with $\tilde{\sigma} ^2$ denoting the noise power on each subcarrier $\tilde{\sigma} ^2=\Delta fN_0=\sigma^2/P$.
The channel between the BS and UE $u$ can then be deduced from \eqref{Res_r_u}, expressed as follows,
\begin{equation}
\abovedisplayshortskip=2pt
\belowdisplayshortskip=2pt
\abovedisplayskip=2pt
\belowdisplayskip=2pt
\begin{aligned}
\boldsymbol{h}_{u,p,q}^{H}=\sum_{l=1}^{L_u}{\alpha _{u,l}\boldsymbol{a}_{u,l}^T e^{j2\pi \left( \tilde{f}_{D,u,l}qT_O-p\Delta f\tilde{\tau}_{u,l} \right)}}.
\end{aligned}
\label{Channel_TF}
\end{equation}
The achievable communication sum rate is then given by
\begin{equation}
\begin{aligned}
C_{\mathrm{sum}}=\frac{1}{QT_O}\sum_{u=0}^{U-1}{\sum_{\left( p,q \right) \in \varGamma _u}{\log _2\left( 1+\frac{\left\| \boldsymbol{h}_{u,p,q}^{H}\boldsymbol{f}_u \right\| ^2}{\sigma ^2} \right)}}.
\end{aligned}
\label{C_sum}
\end{equation}

The received sensing signal can be written as
\begin{equation}
\abovedisplayshortskip=2pt
\belowdisplayshortskip=2pt
\abovedisplayskip=2pt
\belowdisplayskip=2pt
\begin{aligned}
\boldsymbol{y}\left( t \right) =\sum_{k=1}^{K}{\boldsymbol{y}_k\left( t \right)}+\boldsymbol{z}\left( t \right),
\end{aligned}
\label{y}
\end{equation}
in which $\boldsymbol{z}(t)\sim\mathcal{CN} \left( 0,\tilde{\sigma}^2\boldsymbol{I}_{M} \right)$ and the signal reflected from target $k$ can be written as
\begin{equation}
\abovedisplayshortskip=2pt
\belowdisplayshortskip=2pt
\abovedisplayskip=2pt
\belowdisplayskip=2pt
\begin{aligned}
\boldsymbol{y}_k\left( t \right) &=\int_{-\infty}^{\infty}{\boldsymbol{H}_{TG,k}^{T}\left( t,\tau \right) \boldsymbol{s}\left( t-\tau \right)d\tau}
\\
&=\alpha _k\boldsymbol{a}_k \boldsymbol{a}_k^T \boldsymbol{s}\left( t-\tau _k \right) e^{j2\pi f_{D,k}t}.
\end{aligned}
\label{y_k}
\end{equation}
The received data tensor $\mathcal{Y} \in \mathbb{C} ^{MN\times {P}\times {Q}}$ can be expressed as
\begin{equation}
\abovedisplayshortskip=2pt
\belowdisplayshortskip=2pt
\abovedisplayskip=2pt
\belowdisplayskip=2pt
\begin{aligned}
\mathcal{Y} =\sum_{k=0}^{K-1}{\mathcal{Y} _k}+\mathcal{Z},
\end{aligned}
\label{y_tensor}
\end{equation}
where $[\mathcal{Z}]_{:,{p},{q}}\sim\mathcal{CN} \left( 0,\sigma ^2\boldsymbol{I}_{MN} \right)$ and $\mathcal{Y} _k$ is the tensor corresponding to the $k$th target  \cite{dai2025tutorial},
\begin{equation}
\abovedisplayshortskip=2pt
\belowdisplayshortskip=2pt
\abovedisplayskip=2pt
\belowdisplayskip=2pt
\begin{aligned}
\left[ \mathcal{Y} _k \right] _{:,{p},{q}}\!=\!\alpha _k\boldsymbol{a}_k \beta _{k,{p},{q}}e^{-j2\pi  {p}\Delta f\tau _k} e^{j2\pi f_{D,k}{q}T_O} \!\in\! \mathbb{C} ^{MN\times 1},
\end{aligned}
\label{y_k_tensor}
\end{equation}
in which $\left[\cdot\right] _{:,{p},{q}}$ denotes the first dimension of a tensor with the second and third dimension index fixed as ${p}$ and ${q}$. The equivalent modulation symbol $\beta _{k,{p},{q}}$ can be written as
\begin{equation}
\abovedisplayshortskip=2pt
\belowdisplayshortskip=2pt
\abovedisplayskip=2pt
\belowdisplayskip=2pt
\begin{aligned}
\beta _{k,p,q}=\boldsymbol{a}_k^T\boldsymbol{f}_{p,q}b_{p,q},
\end{aligned}
\label{eq_mod_symb}
\end{equation}
with $\boldsymbol{f}_{p,q}$ expressed as
\begin{equation}
\abovedisplayshortskip=2pt
\belowdisplayshortskip=2pt
\abovedisplayskip=2pt
\belowdisplayskip=2pt
\begin{aligned}
\boldsymbol{f}_{p,q}=\begin{cases}
	\boldsymbol{f}_{u,q},\,\,\mathrm{if}\,(p,q)\in \varGamma _u, \forall u = 0,\ldots,U-1,\\
	\boldsymbol{f}_{s,q},\,\,\mathrm{if}\,(p,q)\in \varGamma _s.\\
\end{cases}
\end{aligned}
\label{f_n,q}
\end{equation}

\section{Dedicated versus Zero Sensing Resource Allocation}\label{sec_resource}
\begin{figure}[!htbp]
\vspace*{-8pt}
\centering
\includegraphics[width=0.35\textwidth]{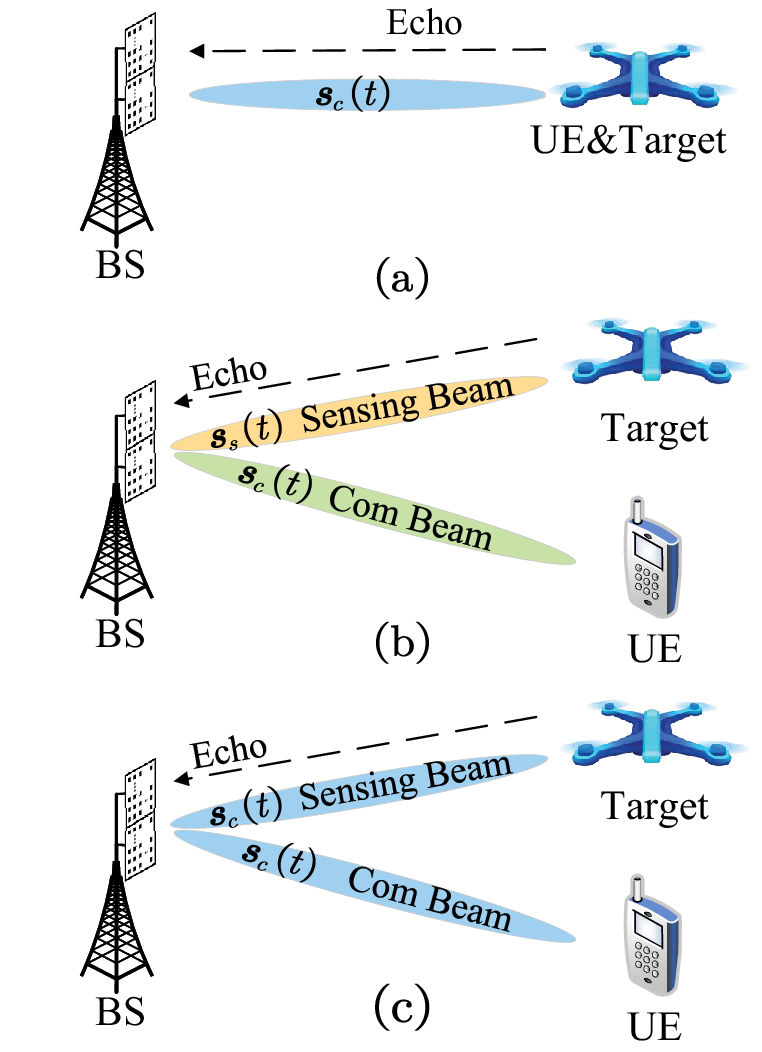}
\vspace*{-6pt}
\caption{ISAC scenarios with (a) shared signal for collocated UE and sensing target, (b) independent signals for separate UE and target; and (c) shared signal for separate UE and target.}
\label{motivation}
\vspace*{-6pt}
\end{figure}

To gain some insights, we first consider a special case illustrated in Fig. \ref{motivation} (a), where the communication UE is also the sensing target \cite{10233570}. In this case, the reflected communication signal can be directly used for sensing, making the dedicated sensing signal $\boldsymbol{s}_s(t)$ unnecessary. On the other hand, in the scenario shown in Fig. \ref{motivation} (b) and (c), where the sensing target and communication UE are distinct, reusing the communication signal for sensing remains feasible, as long as the target is relatively close to the UE in the angular domain so that they can be served by a same beam. However, when the target and UE are far apart, directly reusing the communication signal for sensing could lead to degraded performance, as $\boldsymbol{s}_c(t)$ is beamformed towards the user, which is usually weak at the target's direction. To address this issue, a dedicated sensing signal $\boldsymbol{s}_s(t)$ can be transmitted. To avoid interference between communication and sensing signals, orthogonal resource allocation is often employed, i.e., $\varGamma _s\cap \varGamma _u=\varnothing , \forall u = 1,\ldots,U$. In the following, this approach is used as a benchmark and is referred to as the case with dedicated sensing resources.

\subsection{Dedicated Sensing TF Resource Allocation}
The performance of delay-Doppler sensing with dedicated sensing resources is highly dependent on the resource allocation strategy. In this paper, we consider a simple block-wise resource allocation structure. In this case, the allocated sensing resources $\varGamma _s$ can be expressed as,
\begin{equation}
\begin{aligned}
\varGamma _s\!=\!\{ ( p,q ) |p_s\!\le\! p\!\le\! p_s\!+\!P_s\!-\!1,q_s\!\le\! q\!\le\! q_s\!+\!Q_s\!-\!1 \},
\end{aligned}
\label{Sensing_Resources}
\end{equation}
where $p_s$ and $q_s$ denote the start subcarrier and OFDM symbol indices, respectively, and $P_s$ and $Q_s$ denote the number of sensing subcarriers and OFDM symbols. 
Sensing with dedicated TF resources is simple and straightforward since it requires little modifications to the existing communication system, other than the allocation of designated TF resources for sensing purposes.

However, this scheme results in degraded communication capacity and decreased sensing performance. To better understand this, let's consider a simple example where the BS serves one UE and simultaneously senses one target. For simplicity, assume that the UE is at angle $(\bar{\phi}_U,\bar{\theta}_U)$ with only the line-of-sight (LoS) path. Denote the proportion of dedicated sensing subcarriers and OFDM symboks as $\kappa _P\triangleq P_s/P\in [0,1]$ and $\kappa _Q\triangleq Q_s/Q\in [0,1]$ respectively. The proportion of dedicated sensing can then be expressed as $\kappa = \kappa _P\kappa _Q$. With all the remaining TF resources allocated to the UE, i.e., $\varGamma_U =\varGamma\backslash\varGamma _s$, the achievable communication rate can be expressed as
\begin{equation}
\begin{aligned}
C_U=\frac{\left( 1-\kappa \right) P}{T_O}\log _2\left( 1+\frac{\left\| \alpha _U\boldsymbol{a}_{U}^{T}\boldsymbol{f}_U \right\| ^2}{\sigma ^2} \right),
\end{aligned}
\label{C_U}
\end{equation}
where $\boldsymbol{a}_{U}=\boldsymbol{a}( \phi_U ,\theta_U )$ and $\alpha _U$ is the corresponding complex scattering coefficient.
Note that with only the LoS path, the communication channel after synchronization is frequency and time invariant, i.e., $\boldsymbol{h}_{U,p,q}^{H}=\alpha _U\boldsymbol{a}_{U}^{T},\,\forall \left( p,q \right) \in \varGamma _U$. For dedicated sensing resources, the communication signal is not transmitted on $(p,q)\in\varGamma_s$, the average communication transmit power is,
\begin{equation}
\begin{aligned}
P_{\mathrm{Tx},c}=(1-\kappa) \left\| \boldsymbol{f}_U \right\| ^2.
\end{aligned}
\label{P_Tx,c}
\end{equation}
With maximum ratio transmission (MRT) beamforming, i.e., $\boldsymbol{f}_U=\boldsymbol{a}_{U}^{*}\sqrt{\frac{P_{\mathrm{Tx},c}}{M\left( 1-\kappa \right)}}$, \eqref{C_U} can be further written as
\begin{equation}
\begin{aligned}
C_U=\frac{\left( 1-\kappa \right) P}{T_O}\log _2\left( 1+\frac{MP_{\mathrm{Tx},c}\left| \alpha _U \right|^2}{(1-\kappa )\sigma ^2} \right) .
\end{aligned}
\label{C_U_better}
\end{equation}
Since $\left( 1-\kappa \right)$ is in the pre-log factor, it is easy to prove that $C_U$ is a monotonically decreasing function of $\kappa$. With dedicated sensing resource allocation, $\kappa>0$, and the communication capacity is degraded. 

Besides, with all the sensing TF resources allocated to the target, we can derive the sensing SNR as
\begin{equation}
\begin{aligned}
\mathring{\gamma}_S=\frac{\left| \alpha _k\boldsymbol{a}_{k}^{T}\boldsymbol{f}_S \right|^2}{\sigma ^2}=\frac{\left| \alpha _k\boldsymbol{a}_{k}^{T}\boldsymbol{f}_S \right|^2}{N_0P\Delta f}.
\end{aligned}
\label{SSNR}
\end{equation}
Then, by extending the averaged Cram\'er-Rao lower bound (ACRB) proposed in
\cite{2014ofdmradar} to include averaging in the spatial domain, we can obtain the ACRB for delay and Doppler as
\begin{equation}
\begin{aligned}
\mathrm{ACRB}_{\tau}&=\frac{6}{\left( P_s^2-1 \right) P_sQ_sMN\mathring{\gamma}_S}\cdot \left( \frac{1}{2\pi \Delta f} \right) ^2
\\
&=\frac{3N_0P}{2\pi ^2\left| \alpha _k\boldsymbol{a}_{k}^{T}\boldsymbol{f}_S \right|^2\left( P_s^2-1 \right)P_s Q_sMN\Delta f},
\\
\mathrm{ACRB}_{f_D}&=\frac{6}{\left( Q_{s}^{2}-1 \right) Q_sP_sMN\mathring{\gamma}_S}\cdot \left( \frac{1}{2\pi T_O} \right) ^2
\\
&=\frac{3N_0P\Delta f}{2\pi ^2\left| \alpha _k\boldsymbol{a}_{k}^{T}\boldsymbol{f}_S \right|^2\left( Q_{s}^{2}-1 \right) Q_sP_sMNT_{O}^{2}}.
\end{aligned}
\label{CRLB}
\end{equation}
Note that for dedicated sensing resources, since the sensing signal is only transmitted on $P_s$ subcarriers and $Q_s$ OFDM symbols, the average sensing transmit power is,
\begin{equation}
\begin{aligned}
P_{\mathrm{Tx},s}=\kappa \left\| \boldsymbol{f}_S \right\| ^2=\frac{P_sQ_s}{PQ}\left\| \boldsymbol{f}_S \right\| ^2.
\end{aligned}
\label{P_Tx,s}
\end{equation}
Denote by $\bar{\boldsymbol{f}}_S=\boldsymbol{f}_S/\left\| \boldsymbol{f}_S \right\| $ the normalized beamforming vector, \eqref{CRLB} can be rewritten as
\begin{equation}
\begin{aligned}
\mathrm{ACRB}_{\tau}&=\frac{3N_0}{2\pi ^2P_{\mathrm{Tx},s}\left| \alpha _k\boldsymbol{a}_{k}^{T}\bar{\boldsymbol{f}}_S \right|^2\left( \kappa _{P}^{2}P^2-1 \right) QMN\Delta f},
\\
\mathrm{ACRB}_{f_D}&=\frac{3N_0\Delta f}{2\pi ^2P_{\mathrm{Tx},s}\left| \alpha _k\boldsymbol{a}_{k}^{T}\bar{\boldsymbol{f}}_S \right|^2\left( \kappa _{Q}^{2}Q^2-1 \right) QMNT_{O}^{2}}.
\end{aligned}
\label{CRLB_Clean}
\end{equation}
The delay and Doppler resolution is
\begin{equation}
\begin{aligned}
\Delta \tau =\frac{1}{\kappa _PP\Delta f},
\\
\Delta f_D=\frac{1}{\kappa _QQT_O}.
\end{aligned}
\label{Res}
\end{equation}
Since $\kappa _P \le 1$ and $\kappa _Q \le 1$, the ACRB and resolution for delay and Doppler sensing are degraded compared to using all the TF resources for sensing. 

\subsection{Zero Sensing TF Resource Allocation}
\begin{figure}[!htbp]
\vspace*{-8pt}
\centering
\includegraphics[width=0.4\textwidth]{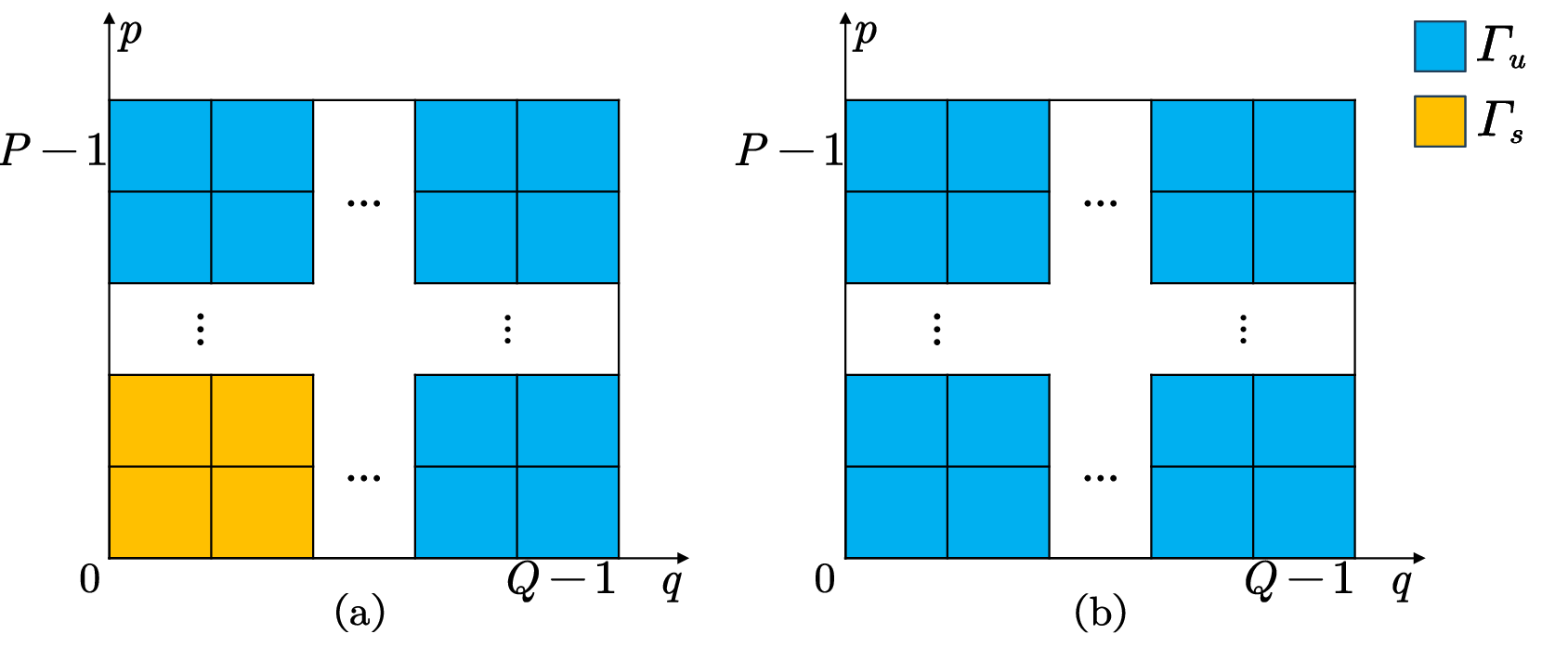}
\vspace*{-6pt}
\caption{OFDM resource grid with (a) dedicated sensing TF resources and (b) zero sensing TF resource.}
\label{ofdmgrid}
\vspace*{-6pt}
\end{figure}

To avoid degradation of communication and sensing performance arising from reduced TF resources, as illustrated in Fig. \ref{motivation} (c), we propose an innovative method with zero sensing TF resource allocation, i.e., $\varGamma_s = \varnothing$ and $\boldsymbol{s}_s\left( t \right)=0$. In this approach, all the TF elements are allocated to communication UEs to maximize communication throughput, while beamforming is designed to meet the requirements for both sensing and communication. The details of the beamforming design will be further discussed in Section \ref{Beamforming Design}.
Fig. \ref{ofdmgrid} illustrates an example of the OFDM resource grid, comparing the case with dedicated sensing TF resources to our proposed method with zero sensing TF resource. In Fig. \ref{ofdmgrid} (b), the TF resources previously dedicated to sensing in Fig. \ref{ofdmgrid} (a) can be repurposed for communication, thereby increasing the spectral efficiency.

\section{Beamforming Design}  \label{Beamforming Design}
\begin{figure}[!htbp]
\vspace{-6pt}
\centering
\includegraphics[width=0.5\textwidth]{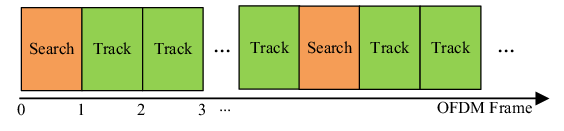}
\vspace*{-6pt}
\caption{Illustration of the sensing procedures at the BS.}
\label{SearchingandTracking}
\end{figure}

As shown in Fig.\ref{SearchingandTracking}, the sensing procedures at the BS can be divided into two main categories: target searching and target tracking. For target searching, the BS performs beam sweeping to identify targets. For target tracking, the BS continuously monitors the identified targets, adjusting its beamforming direction based on predicted angles and updating the parameter estimates for each target. Target searching is carried out periodically to identify new targets, while target tracking occurs in between searching intervals to maintain continuous tracking of the identified targets.

During target searching, since beam sweeping is performed, the transmit beamforming is designed for each sensing beam and each UE. Since no prior knowledge of the targets are available at this stage, the design objective is to minimize the transmit power while ensuring sufficient sensing power at the sensing beam direction and enough communication SNR for the user. 
During target tracking, since prior information of the targets are obtained with previous searching and tracking procedures, beam sweeping is no longer needed and the transmit beamforming is designed for each user. With the scattering coefficients $\{\alpha_k\}_{k=1}^{K}$ estimated,  the design objective is to minimize the transmit power while ensuring enough sensing SNR for each target of interest and communication SNR for the user.

\subsection{Target Searching}
In the target searching stage, the BS adjusts its beamforming to different angles across various OFDM symbols to detect targets. 
The sweeping beam angles can be designed as follows,
\begin{equation}
\begin{aligned}
&\phi _{s,q}=\mathrm{arccos} \left[ \frac{\lambda}{d}\left( \left. \frac{1}{2}-\frac{1}{M}\left. \left\lfloor \frac{M\tilde{q}}{\tilde{Q}} \right. \right. \right\rfloor \right) \right],
\\
&\theta_{s,q}=\mathrm{arccos}\left[ \frac{\lambda}{d}\left( \frac{1}{2}-\frac{1}{N}\mathrm{mod}\left( \left\lfloor \frac{\tilde{q}MN}{\tilde{Q}} \right\rfloor ,N \right) \right) \right],
\end{aligned}
\label{max_range_scheme}
\end{equation}
where $(\phi _{s,q},\theta _{s,q})$ is the sensing beam angle for the $q$th symbol, $\tilde{q}=q, q=0,\ldots,Q-1$ for our proposed method and $\tilde{q}=q-q_s, q=q_s,\ldots,q_s+Q_s-1$ for dedicated sensing resources. The beamforming design for covering these sensing beam directions will be discussed later. Note that with our proposed method, $\tilde{Q}=Q$, while with dedicated sensing resources, $\tilde{Q}=Q_s$. In this design, the OFDM symbols with the same beam direction form a block-like structure. For example, symbols $\tilde{q}=0,1,\ldots,\tilde{Q}_b-1 $ correspond to the same beam angle $(\mathrm{arccos} \left( \frac{\lambda}{2d} \right) ,\mathrm{arccos} \left( \frac{\lambda}{2d} \right) )$, where $\tilde{Q}_b=\frac{\tilde{Q}}{MN}$. 

The design of $Q$ and $Q_s$ should be larger than the number of beams $MN$ to ensure full coverage. Moreover, $Q$ and $Q_s$ should be integer multiples of $MN$ to maintain an equal number of symbols for each beam. 
\subsubsection{Target Searching with Dedicated Sensing Resources}
When dedicated sensing resources are allocated, the beamforming design is straightforward, i.e., 
\begin{equation}
\abovedisplayshortskip=2pt
\belowdisplayshortskip=2pt
\abovedisplayskip=2pt
\belowdisplayskip=2pt
\begin{aligned}
\boldsymbol{f}_{s,q}=\boldsymbol{a}_{s,q}^*,
\end{aligned}
\label{simple_mrc}
\end{equation}
where $\boldsymbol{a}_{s,q}\triangleq\boldsymbol{a}\left( \phi _{s,q},\theta _{s,q} \right)$. In this case, the beamforming vectors can be drawn from a discrete fourier transform (DFT) codebook. 


\subsubsection{Target Searching with Zero Sensing Resource Allocation}
For our proposed method with zero sensing TF resource allocation, the beamforming vector is designed to ensure communication quality with the users while providing sufficient sensing power for target searching.
It is important to note that the channel expressed in \eqref{Channel_TF} is TF dependent, meaning it varies with the subcarrier index $p$ and OFDM symbol index $q$. Designing beamforming directly using this channel would incur significant channel estimation overhead, as it requires knowledge of the channel across all subcarriers and OFDM symbols. Additionally, designing a different beamforming vector for each subcarrier and symbol would be computationally prohibitive and impractical for real-world implementation. Instead, in the following, we design our beamforming based on the statistical channel, which can be calculated as the eigenvector corresponding to the largest eigenvalue of the channel auto-correlation matrix,
\begin{equation}
\abovedisplayshortskip=2pt
\belowdisplayshortskip=2pt
\abovedisplayskip=2pt
\belowdisplayskip=2pt
\begin{aligned}
\boldsymbol{R}_u=\frac{1}{\left| \varGamma _u \right|}\sum_{\left( p,q \right) \in \varGamma _u}{\boldsymbol{h}_{u,p,q}\boldsymbol{h}_{u,p,q}^{H}}.
\end{aligned}
\label{Channel_covariance}
\end{equation}
Let $\lambda_u^2$ denote the largest eigenvalue of $\boldsymbol{R}_u$ and $\boldsymbol{h}_u$ the corresponding eigenvector. The beamforming vector for the $u$th UE can be designed by solving the following optimization problem:
\begin{equation}
\begin{aligned}
&\underset{\boldsymbol{f}_{u,q}}{\min}\,\,\left\| \boldsymbol{f}_{u,q} \right\|^2 
\\
&\mathrm{s}.\mathrm{t}.\,\frac{\left| \lambda _u\boldsymbol{h}_{u}^{H}\boldsymbol{f}_{u,q} \right|^2}{\sigma ^2}\ge \bar{\gamma}\,,
\\
&\,\,\,\,\,\,\,\,\,\frac{\left| \boldsymbol{a}_{s,q}^T \boldsymbol{f}_{u,q} \right|^2}{MN}\ge \mathcal{E} _{\min}\,,
\end{aligned}
\label{Searching_Beamforming}
\end{equation}
where $\bar{\gamma}$ is the minimum communication SNR and $\mathcal{E} _{\min}$ is the minimum sensing power. 

The above formulation is a classic quadratically constrained quadratic programming (QCQP) problem, which can be addressed using semi-definite relaxation (SDR) techniques. Specifically, the first constraint can be rewritten as ${\lambda _u}^2\left( \boldsymbol{h}_{u}^{H}\boldsymbol{f}_{u,q}\boldsymbol{f}_{u,q}^{H}\boldsymbol{h}_u \right) \ge \sigma ^2\bar{\gamma}$. We further introduce the variable $\boldsymbol{F}_{u,q}\triangleq \boldsymbol{f}_{u,q}\boldsymbol{f}_{u,q}^{H}$, which allows us to rewrite the constraint as ${\lambda _u}^2\mathrm{tr}\left( \boldsymbol{h}_u\boldsymbol{h}_{u}^{H}\boldsymbol{F}_{u,q} \right) \ge \sigma ^2\bar{\gamma}$. Similarly, the second constraint becomes $\mathrm{tr}\left( \boldsymbol{a}_{s,q}^* \boldsymbol{a}_{s,q}^T \boldsymbol{F}_{u,q} \right) \ge MN\varepsilon _{\min}$, and the optimization problem \eqref{Searching_Beamforming} can be converted into
\begin{equation}
\begin{aligned}
&\underset{\boldsymbol{F}_{u,q}}{\mathrm{min}}\,\,\mathrm{tr}\left\| \boldsymbol{F}_{u,q} \right\| 
\\
&\mathrm{s}.\mathrm{t}.\,{\lambda _u}^2\mathrm{tr}\left( \boldsymbol{h}_u\boldsymbol{h}_{u}^{H}\boldsymbol{F}_{u,q} \right) \ge \sigma ^2\bar{\gamma},
\\
&\,\,\,\,\,\,\,\,\,\mathrm{tr}\left( \boldsymbol{a}_{s,q}^* \boldsymbol{a}_{s,q}^T \boldsymbol{F}_{u,q} \right) \ge MN\mathcal{E} _{\min},
\\
&\,\,\,\,\,\,\,\,\,\boldsymbol{F}_{u,q}\succeq 0,
\\
&\,\,\,\,\,\,\,\,\,\mathrm{rank}\left( \boldsymbol{F}_{u,q} \right) =1.
\end{aligned}
\label{Searching_Beamforming_SDP}
\end{equation}
The above problem is non-convex due to the rank-1 constraint imposed on $\boldsymbol{F}_{u,q}$. By relaxing this constraint, we can transform the problem into a convex semi-definite programming (SDP) problem. Since problem \eqref{Searching_Beamforming} is a complex-valued homogeneous QCQP with no more than 3 constraints, the solution to the relaxed problem is rank-1 and thus optimal for the original problem \eqref{Searching_Beamforming} \cite{SDR}.


\begin{figure}[!htbp]
\centering
\subfloat[Proposed SDR beamforming]{
\includegraphics[width=\linewidth]{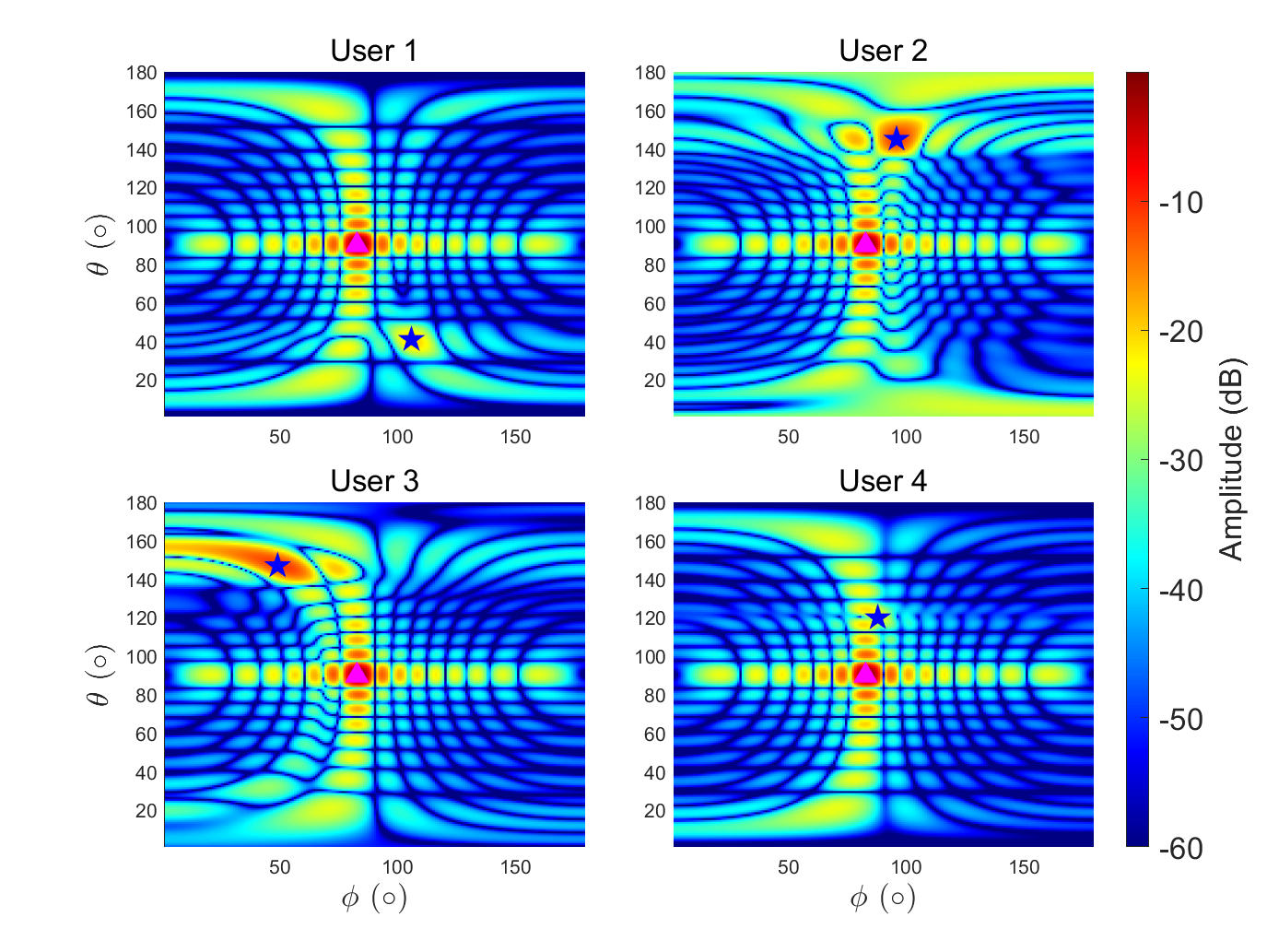}
\label{bp_proposed}
}
\hfill
\subfloat[Conventional MRT beamforming]{
\includegraphics[width=\linewidth]{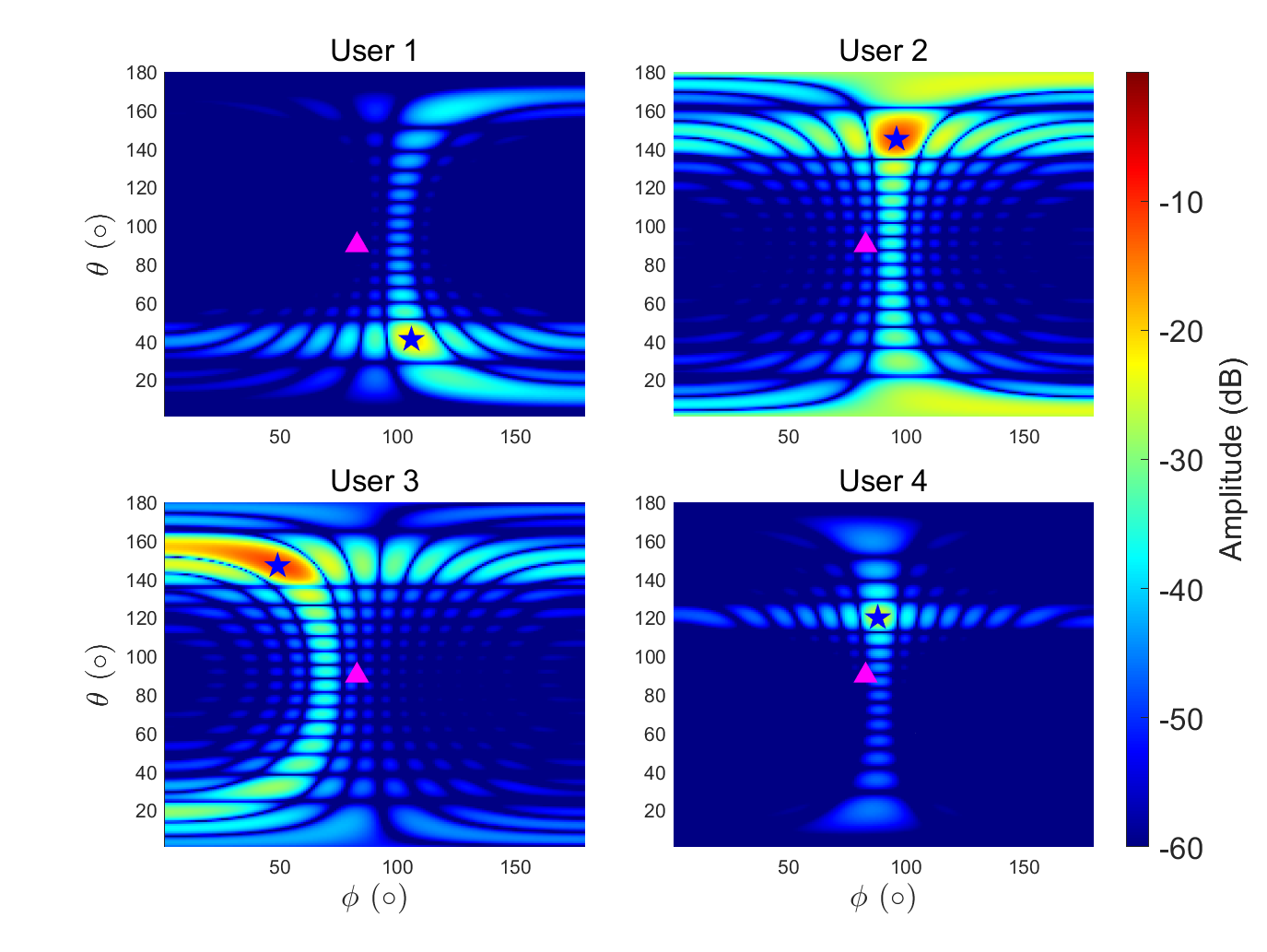}
\label{bp_mrt}
}
\caption{Beam patterns of (a) 4 users sharing TF resources with sensing beam at direction $(82.82, 90)^\circ$ using proposed SDR beamforming and (b) 4 users using conventional MRT beamforming, with user directions marked by blue stars and sensing beam direction marked by pink triangles.}
\label{comparative_beam_patterns}
\vspace*{-8pt}
\end{figure}

Fig. \ref{bp_proposed} illustrates the beam patterns of four users using our proposed SDR-based transmit beamforming method, showing simultaneous coverage for the communication users and a dedicated sensing beam directed at $(82.82, 90)^\circ$. In contrast, Fig. \ref{bp_mrt} presents the conventional MRT beamforming patterns optimized solely for communication channels. A comparison of the two beam patterns reveals that our proposed method achieves dual-function radiation by allocating additional power toward the sensing beam direction while maintaining communication quality.

The above optimization can be computationally demanding since the beamforming vectors for $MN$ beams need to be computed. Next, we propose a low-complexity, closed-form sub-optimal beamforming, expressed as the linear combination of $\boldsymbol{h}_u$ and $\boldsymbol{a}_{s,q}^*$,
\begin{equation}
\begin{aligned}
\boldsymbol{f}_{u,q}=a_u\boldsymbol{h}_u+b_{u,q}e^{j\varphi _{u,q}}\boldsymbol{a}_{s,q}^{*},
\end{aligned}
\label{Closed-Beamforming}
\end{equation}
where $a_u \in \mathbb{R}$ and $b_{u,q} \in \mathbb{R}$ are real coefficients and $\varphi _{u,q}$ is the phase adjustment to ensure constructive addition.
With the beamforming in \eqref{Closed-Beamforming}, the communication SNR can be expressed as
\begin{equation}
\begin{aligned}
\gamma \,=\frac{\left| \lambda _u\boldsymbol{h}_{u}^{H}\boldsymbol{f}_{u,q} \right|^2}{\sigma ^2}\,=\frac{\left| \lambda _ua_u+\lambda _ub_{u,q}e^{j\varphi _{u,q}}\boldsymbol{h}_{u}^{H}\boldsymbol{a}_{s,q}^{*} \right|^2}{\sigma ^2}.
\end{aligned}
\label{Closed-Beamforming-comsnr}
\end{equation}
First, we design $a_u$ to satisfy the minimum communication SNR constraint,
\begin{equation}
\begin{aligned}
a_u=\sqrt{\bar{\gamma}}\frac{\sigma}{\lambda _u}.
\end{aligned}
\label{Closed-Beamforming—coeff1}
\end{equation}
Next, the phase $\varphi _{u,q}$  is designed to ensure that the communication SNR $\gamma $ does not decrease with the addition of the second term in \eqref{Closed-Beamforming}. In order to achieve this, $e^{j\varphi _{u,q}}\boldsymbol{h}_{u}^{H}\boldsymbol{a}_{s,q}^{*}$ is designed to be a positive real number with
\begin{equation}
\begin{aligned}
\varphi _{u,q} = \angle \left( \boldsymbol{a}_{s,q}^{T}\boldsymbol{h}_u \right).
\end{aligned}
\label{Closed-Beamforming—phase}
\end{equation}
Finally, we adjust $b_{u,q}$ to satisfy the sensing power constraint. With the closed-form beamforming, the sensing power can be expressed as,
\begin{equation}
\begin{aligned}
\mathcal{E} =\frac{\left| \boldsymbol{a}_{s,q}^{T}\boldsymbol{f}_{u,q} \right|^2}{MN}=\,\frac{\left( \left| \sqrt{\bar{\gamma}}\frac{\sigma}{\lambda _u}\boldsymbol{a}_{s,q}^{T}\boldsymbol{h}_u \right|+\left| MNb_{u,q} \right| \right) ^2}{MN}.
\end{aligned}
\label{Closed-Beamforming-sensepwr}
\end{equation}
Thus, the minimum $b_{u,q}$ required to satisfy the second constraint is
\begin{equation}
\begin{aligned}
b_{u,q}=\max \left\{ \frac{1}{MN}\left( \sqrt{MN\mathcal{E} _{\min}}-\left| \sqrt{\bar{\gamma}}\frac{\sigma}{\lambda _u}\boldsymbol{a}_{s,q}^{T}\boldsymbol{h}_u \right| \right) ,0 \right\}.
\end{aligned}
\label{Closed-Beamforming-coeff2}
\end{equation}

\subsection{Target Tracking} \label{target_tracking}
In the target tracking stage, the BS utilizes multiple beams to effectively cover the targets of interest. The beam angles employed in this stage are derived from previous target searching or tracking results. Note that since beam sweeping is not needed in this stage, the beamforming vectors $\boldsymbol{f}_{s,q}$ and $\boldsymbol{f}_{u,q}$ are constant and independent of $q$. Thus, we omit the subscript $q$ in the following discussions.
\subsubsection{Target Tracking with Dedicated Sensing Resources} \label{Tracking_dedicated}
For target tracking utilizing dedicated sensing resources, the beamforming strategy is to provide enough sensing SNR for all the targets of interest. 
Denote by $\varOmega_{\rm{trk}}$ the set of targets of interest. For each $k\in\varOmega_{\rm{trk}}$, let ($\tilde{\phi}_{k},\tilde{\theta}_{k}$) and $\tilde{\alpha}_k$ represent the AoA and complex coefficient estimate from prior target searching or tracking. The beamforming vector can then be obtained by solving the optimization problem below,
\begin{equation}
\begin{aligned}
&\underset{\boldsymbol{f}_s}{\mathrm{min}}\,\,\left\| \boldsymbol{f}_s \right\| ^2
\\
&\mathrm{s.t.}
\frac{\left| \tilde{\alpha}_k\tilde{\boldsymbol{a}}_k^T \boldsymbol{f}_s \right|^2}{\sigma^2}\ge \bar{\gamma}_s,\forall k\in \varOmega _{\mathrm{trk}},
\end{aligned}
\label{Tracking_Beamforming}
\end{equation}
where $\tilde{\boldsymbol{a}}_k\triangleq \boldsymbol{a}\left( \tilde{\phi}_k,\tilde{\theta}_k \right) $ and $\bar{\gamma}_s$ is the minimum required sensing SNR. Let $\boldsymbol{F}_s\triangleq \boldsymbol{f}_s\boldsymbol{f}_{s}^{H}$, this problem can be solved with SDR,
\begin{equation}
\begin{aligned}
&\underset{\boldsymbol{F}_s}{\mathrm{min}}\,\,\mathrm{tr}\left\| \boldsymbol{F}_s \right\| 
\\
&\mathrm{s}.\mathrm{t}.\,\tilde{\alpha}_{k}^{2}\mathrm{tr}\left( \tilde{\boldsymbol{a}}_k^* \tilde{\boldsymbol{a}}_k^T \boldsymbol{F}_s \right) \ge {\sigma}^2\bar{\gamma}_s,\forall k\in \varOmega _{\mathrm{trk}},
\\
&\,\,\,\,\,\,\,\,\,\boldsymbol{F}_s\succeq 0.
\end{aligned}
\label{Tracking_Beamforming_SDR}
\end{equation}
When the number of targets of interest $\left| \varOmega \right|\le 3$, the solution is tight and problem \eqref{Tracking_Beamforming} can be solved optimally. For $\left| \varOmega \right| > 3$, we employ Gaussian randomization to obtain an approximate solution \cite{SDR}. 

\subsubsection{Target Tracking with Zero Sensing Resource Allocation} \label{Tracking_proposed}
For our proposed method, in addition to tracking targets, the BS is required to ensure sufficient signal for communication UEs. For each UE $u$, the optimization problem is formulated as follows,
\begin{equation}
\begin{aligned}
&\mathrm{min}\,\,\left\| \boldsymbol{f}_u \right\|^2
\\
&\mathrm{s}.\mathrm{t}.\frac{\left| \lambda _u\boldsymbol{h}_{u}^{H}\boldsymbol{f}_u \right|^2}{\sigma ^2}\ge \bar{\gamma}\,,
\\
&\,\,\,\,\,\,\,\,\,\frac{\left| \tilde{\alpha}_k\tilde{\boldsymbol{a}}_k^T \boldsymbol{f}_u \right|^2}{\sigma^2}\ge \bar{\gamma}_s,\forall k\in \varOmega _{\mathrm{trk}}.
\end{aligned}
\label{Tracking_Beamforming_Proposed}
\end{equation}
In contrast to problem \eqref{Tracking_Beamforming}, this formulation incorporates a communication SNR constraint to ensure adequate communication performance for the users. As a result, it can be solved exactly with SDR when $\left| \varOmega \right|\le 2$. When $\left| \varOmega \right|> 2$, Gaussian randomization techniques are employed.

\section{Sensing Algorithm} \label{Sensing Algo}

\begin{figure}[!htbp]
\vspace{-6pt}
\centering
\includegraphics[width=0.45\textwidth]{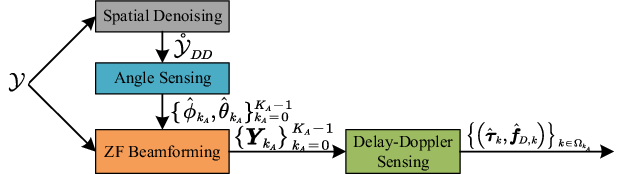}
\vspace*{-6pt}
\caption{Signal processing procedure for sensing.}
\label{flowgraph}
\end{figure}

Fig. \ref{flowgraph} illustrates the overall signal processing procedure for sensing. First, a novel denoising method is proposed to enhance the SNR without disrupting the steering vector structure in the spatial domain. Next, the MUSIC algorithm is used to estimate the AoAs of the targets, which are then utilized for zero-forcing (ZF) beamforming to separate the signals from each path. Finally, delay-Doppler sensing is employed to estimate the delays and Doppler shifts of each target.

\subsection{Spatial Domain Denoising} \label{Denoise}
In this study, MUSIC algorithm is utilized for AoA estimation due to its high-resolution. It is preferable to apply MUSIC following delay-Doppler processing to harness the substantial processing gain. However, for target searching and tracking with our proposed zero sensing resource allocation, the AoAs need to be estimated first to recover the equivalent modulation symbols for element-wise division. This process will be further elaborated in Section \ref{DD_processing}. Moreover, for target searching with dedicated sensing resources, applying MUSIC to all beams together is more efficient than processing each beam separately, as this approach avoids the need to match parameters across beams and provides more snapshots for the MUSIC algorithm. As a result, in order to improve the MUSIC performance, a spatial domain denoising method is proposed. First, based on the resulting sensing signal tensor \eqref{y_tensor}, element-wise division is conducted to remove the randomness of the modulation symbols:
\begin{equation}
\begin{aligned}
&[ \tilde{\mathcal{Y}} ] _{:,p,q}=\frac{\left[ \mathcal{Y} \right] _{:,p,q}}{b_{p,q}}
\\
&=\sum_{k=0}^{K-1}{\alpha _k\boldsymbol{a}_k\boldsymbol{a}_k^T\boldsymbol{f}_{p,q}e^{-j2\pi p\Delta f\tau _k}e^{j2\pi f_{D,k}qT_O}}+[ \tilde{\mathcal{Z}} ] _{:,p,q},
\end{aligned}
\label{Denoise_Symbols}
\end{equation}
where $[ \tilde{\mathcal{Z}} ] _{:,p,q}=\left[ \mathcal{Z} \right] _{:,p,q}/b_{p,q}$. Note that \eqref{Denoise_Symbols} can not be utilized directly for delay-Doppler processing due to the presence of the term $\boldsymbol{f}_{p,q}$. However, the signal \eqref{Denoise_Symbols} still exhibits certain sparsity in the delay-Doppler domain, which can be leveraged for denoising in the spatial domain. Thus, we transform the signal \eqref{Denoise_Symbols} to delay-Doppler domain.
For target searching, as beam sweeping is performed, FFT is performed for the symbols of each beam separately,
\begin{equation}
\begin{aligned}
\relax[\tilde{\mathcal{Y}}_{DD}]_{:,g,n_bQ_b+w}=\sum_{\left( p,q \right) \in \tilde{\varGamma}_{s,n_b}}{[\tilde{\mathcal{Y}}]_{:,p,q}e^{j\frac{2\pi}{P}pg}e^{-j\frac{2\pi}{Q}qw}},
\end{aligned}
\label{DD_signal_searching}
\end{equation}
where $n_b = 0,\ldots, MN-1$ is the beam index, $w = 0,\ldots,Q_b-1$, $g = 0, \ldots, P-1$ and $\tilde{\varGamma}_{s,n_b}=\tilde{\varGamma}_s\cap \left\{ \left( p,q \right) ,n_bQ_b\le \tilde{q}<\left( n_b+1 \right) Q_b \right\}$. For dedicated sensing resources, $\tilde{\varGamma}_s=\varGamma_s$  and for our proposed method, $\tilde{\varGamma}_s=\varGamma$.

For target tracking, since no beam sweeping is required, FFT is performed only once across all the sensing symbols,
\begin{equation}
\begin{aligned}
\relax[\tilde{\mathcal{Y}}_{DD}]_{:,g,w}=\sum_{\left( p,q \right) \in \tilde{\varGamma}_s}{[\tilde{\mathcal{Y}}]_{:,p,q}e^{j\frac{2\pi}{P}pg}e^{-j\frac{2\pi}{Q}qw}}.
\end{aligned}
\label{DD_signal}
\end{equation}
Then, the delay-Doppler bins with energy below the threshold $\mathcal{E}_{th}$ are set to zero, yielding the denoised delay-Doppler domain signal,
\begin{equation}
\begin{aligned}
\relax[ \mathring{\mathcal{Y}}_{DD} ] _{:,g,w}=\begin{cases}
	[ \tilde{\mathcal{Y}}_{DD} ] _{:,g,w},\left\| [ \tilde{\mathcal{Y}}_{DD} ] _{:,g,w} \right\| ^2\ge \mathcal{E} _{th}\\
	0\;\;\;\;\;\;\;\;\;\;\;\;\;\;,\mathrm{otherwise}.\\
\end{cases}
\end{aligned}
\label{DD_Denoise}
\end{equation}
Since most noise falls below the threshold $\mathcal{E} _{th}$, this operation enhances the SNR by preserving only those DD domain points whose energy exceeds this level.
\subsection{Angle Sensing}


The angle sensing algorithm is the same for target searching or tracking with either dedicated sensing resources or our proposed method.  The angles $\{\phi_k,\theta_k\}_{k=1}^{K}$ can be estimated using classic spatial spectrum estimation algorithms such as the periodogram method or the MUSIC algorithm. Here, MUSIC is employed for its high resolution.
Forward-backward spatial smoothing techniques are applied to address the potential correlation among signals received from different paths \cite{spatialsmoothing}.
For a 2D planar array, spatial smoothing is performed on both dimensions. The spatial auto-correlation matrix of the $(i,j)$th subarray can be calculated as
\begin{equation}
\abovedisplayshortskip=2pt
\belowdisplayshortskip=2pt
\abovedisplayskip=2pt
\belowdisplayskip=2pt
\begin{aligned}
&\boldsymbol{R}_{i,j}=\frac{1}{PQ}\sum_{g=0}^{P-1}{\sum_{w=0}^{Q-1}{[\mathring{\mathcal{Y}}_{DD}]_{\boldsymbol{c}_{i,j},g,w}[\mathring{\mathcal{Y}}_{DD}]_{\boldsymbol{c}_{i,j},g,w}^{H}}}
\\
&\in \mathbb{C} ^{M_{\mathrm{sub}} N_{\mathrm{sub}}\times M_{\mathrm{sub}} N_{\mathrm{sub}}},
\end{aligned}
\label{Correlation}
\end{equation}
where $i=0,...,I-1$ and $j=0,...,J-1$ with $I$ and $J$ denoting the number of subarrays along each axis. The size of the subarrays is $M_{\mathrm{sub}} \times N_{\mathrm{sub}}$, where $M_{\mathrm{sub}}=M-I+1$ and $N_{\mathrm{sub}}=N-J+1$. Vector $\boldsymbol{c}_{i,j}\in\mathbb{N}^{M_{\mathrm{sub}}N_{\mathrm{sub}}} \times 1$ contains the $(i,j)$th subarray's element indices in the original array steering vector expressed in \eqref{SteeringVec2D}. 
The auto-correlation matrix after 2D spatial smoothing can then be expressed as
\begin{equation}
\abovedisplayshortskip=2pt
\belowdisplayshortskip=2pt
\abovedisplayskip=2pt
\belowdisplayskip=2pt
\begin{aligned}
\boldsymbol{R}_{\mathcal{Y}}=\frac{1}{IJ}\sum_{i=0}^{I-1}{\sum_{j=0}^{J-1}{\boldsymbol{R}_{i,j}}}.
\end{aligned}
\label{Correlation_Smooth}
\end{equation}
The auto-correlation matrix after forward-backward smoothing can be obtained by
\begin{equation}
\abovedisplayshortskip=2pt
\belowdisplayshortskip=2pt
\abovedisplayskip=2pt
\belowdisplayskip=2pt
\begin{aligned}
\boldsymbol{R}_{\mathcal{Y} ,\mathrm{FB}}=\frac{1}{2}\left( \boldsymbol{R}_{\mathcal{Y}}+\boldsymbol{JR}_{\mathcal{Y}}\boldsymbol{J} \right),
\end{aligned}
\label{BF Smoothing}
\end{equation}
where $\boldsymbol{J}$ is the exchange matrix.

Then, eigenvalue decomposition (EVD) is performed on $\boldsymbol{R}_{\mathcal{Y} ,\mathrm{FB}}$ as $\boldsymbol{R}_{\mathcal{Y} ,\mathrm{FB}}=\boldsymbol{E}_s\boldsymbol{\varLambda }_s\boldsymbol{E}_{s}^{H}+\boldsymbol{E}_n\boldsymbol{\varLambda }_n\boldsymbol{E}_{n}^{H}$, where $\boldsymbol{\varLambda}_s$ and $\boldsymbol{\varLambda}_n$ denote the diagonal matrices consisting of $K_A$ large eigenvalues and $M_{\mathrm{sub}} N_{\mathrm{sub}}-K_A$ small eigenvalues respectively, and the columns of $\boldsymbol{E}_s$ and $\boldsymbol{E}_n$ are the corresponding eigenvectors. Then, the MUSIC spectrum is calculated as
\begin{equation}
\begin{aligned}
P_{\mathrm{MUSIC}}\left( \phi,\theta \right) =\frac{1}{\boldsymbol{a}^H\left( \phi,\theta \right) \boldsymbol{E}_n\boldsymbol{E}_{n}^{H}\boldsymbol{a}\left( \phi,\theta \right)}.
\end{aligned}
\label{MUSIC}
\end{equation}
The angles of the sensing targets $\{\hat{\phi}_{k_A},\hat{\theta}_{k_A}\}_{k_A=1}^{K_A}$ can then be estimated by searching the spectrum. Note that $K_A$ is also the number of estimated AoAs. Since the AoAs of different targets may be equal or very close to each other, we have $K_A\le K$.
\subsection{Delay-Doppler Sensing} \label{DD_processing}
For delay-Doppler sensing, the resource grids corresponding to different angles $\{\hat{\phi}_{k_A},\hat{\theta}_{k_A}\}_{k_A=0}^{K_A-1}$ are extracted,
\begin{equation}
\begin{aligned}
\boldsymbol{Y}_{k_A}=\boldsymbol{f}_{k_A}^H\mathcal{Y} \in \mathbb{C} ^{{P}\times {Q}},
\end{aligned}
\label{Signal extraction}
\end{equation}
where $\boldsymbol{f}_{k_A}$ is the ZF beamforming vector towards AoA $(\hat{\phi}_{k_A},\hat{\theta}_{k_A})$. Let $\hat{\boldsymbol{a}}_{k_A}\triangleq \boldsymbol{a}(\hat{\phi}_{k_A},\hat{\theta}_{k_A})$ and $\boldsymbol{A}_{k_A}\triangleq [\hat{\boldsymbol{a}}_1,\!...,\hat{\boldsymbol{a}}_{k_A-1},\hat{\boldsymbol{a}}_{k_A+1},\!...,\hat{\boldsymbol{a}}_{K_A}]$ and denote by $\boldsymbol{Q}_{k_A}\triangleq \boldsymbol{I}_M-\boldsymbol{A}_{k_A}\left( \boldsymbol{A}_{k_A}^{H}\boldsymbol{A}_{k_A} \right) ^{-1}\boldsymbol{A}_{k_A}^{H}$ the projection matrix into the null space of $\boldsymbol{A}_{k_A}^{H}$. Then the ZF beamforming vector can be expressed as $\boldsymbol{f}_{k_A}=\boldsymbol{Q}_{k_A}\hat{\boldsymbol{a}}_{k_A} /\| \boldsymbol{Q}_{k_A}\hat{\boldsymbol{a}}_{k_A} \| $. 
By substituting \eqref{y_tensor} and \eqref{y_k_tensor} into \eqref{Signal extraction}, it can be further expressed as \eqref{TFMatrix}, in which $\varOmega _{k_A}\triangleq \left\{ k|(\phi _k,\theta_k)=(\hat{\phi}_{k_A},\hat{\theta}_{k_A}) \right\} $ denotes the set of targets with AoAs equal to $(\hat{\phi}_{k_A},\hat{\theta}_{k_A})$. In the following analysis, we assume that the AoAs are estimated accurately. Then, the second term in \eqref{TFMatrix} becomes zero due to the design of the ZF beamforming. Specifically, it holds that $\boldsymbol{f}_{k_{\phi ,1}}^{H}\hat{\boldsymbol{a}}_{k_{\phi,2}} =0,\forall k_{\phi ,1}\ne k_{\phi ,2}$.
\begin{figure*}
\begin{equation}
\abovedisplayshortskip=2pt
\belowdisplayshortskip=2pt
\abovedisplayskip=2pt
\belowdisplayskip=2pt
\begin{aligned}
&\left[ \boldsymbol{Y}_{k_A} \right] _{{p},{q}}=\boldsymbol{f}_{k_A}^H\left[ \mathcal{Y} _k \right] _{:,{p},{q}}=\boldsymbol{f}_{k_A}^H\sum_{k=0}^{K-1}{\alpha _k\boldsymbol{a}_k \beta _{k,{p},{q}}e^{-j2\pi {p}\Delta f\tau _k}e^{j2\pi f_{D,k}{q}T_O}}+\left[ \boldsymbol{Z}_{k_A} \right] _{{p},{q}}
\\
&=\underbrace{\sum_{k\in \varOmega _{k_A}}{\alpha _k\boldsymbol{f}_{k_A}^H\hat{\boldsymbol{a}}_{k_A} \beta _{k,{p},{q}}e^{-j2\pi {p}\Delta f\tau _k}e^{j2\pi f_{D,k}{q}T_O}}}_{\mathrm{desiered \,signal}}+\underbrace{\sum_{k\notin \varOmega _{k_A}}{\alpha _k\boldsymbol{f}_{k_A}^H\boldsymbol{a}_k \beta _{k,{p},{q}}e^{-j2\pi {p}\Delta f\tau _k}e^{j2\pi f_{D,k}{q}T_O}}}_{\mathrm{inter-angle \,interference}}+\left[ \boldsymbol{Z}_{k_A} \right] _{{p},{q}}
\end{aligned}
\label{TFMatrix}
\end{equation}
\end{figure*}
Next, the TF signal from each angle $\{ \boldsymbol{Y}_{k_A}\}_{k_A=0}^{K_A-1}$ are processed independently.
\subsubsection{Target Searching with Dedicated Sensing Resources}
For target searching, the symbols corresponding to the closest beam angle are used for delay and Doppler estimation. Specifically, the set of symbols used for targets in $\varOmega _{k_A}$ can be written as
\begin{equation}
\begin{aligned}
\varUpsilon _{k_A}\triangleq &\left\{ q\left| -1/M\le \sin \left( \phi _{s,q} \right) -\sin \left( \hat{\phi}_{k_A} \right) <1/M, \right. \right. 
\\
&\hspace{16pt}\left. -1/M\le \sin \left( \theta _{s,q} \right) -\sin \left( \hat{\theta}_{k_A} \right) <1/M \right\} .
\end{aligned}
\label{Searching_Sensing_Symbol_Set}
\end{equation}
Then, the set of TF resources utilized for target searching to cover angle $(\hat{\phi}_{k_A},\hat{\theta}_{k_A})$ can be expressed as
\begin{equation}
\begin{aligned}
\varGamma _{s,k_A}\triangleq \left\{ \left( p,q \right) |\left( p,q \right) \in \varGamma _s,q\in \varUpsilon _{k_A} \right\}.
\end{aligned}
\label{Searching_Sensing_Resource_Set}
\end{equation}
First, element-wise division is used to remove the effects of modulation symbols and obtain the TF domain channel,
\begin{equation}
\begin{aligned}
&\left[ \tilde{\boldsymbol{Y}}_{s,k_A} \right] _{p,q}=\frac{\left[ \boldsymbol{Y}_{k_A} \right] _{p,q}}{b_{p,q}}
\\
&=\boldsymbol{f}_{k_A}^{H}\hat{\boldsymbol{a}}_{k_A}\hat{\boldsymbol{a}}_{k_A}^T\boldsymbol{f}_{s,k_A}\sum_{k\in \varOmega _{k_A}}{\alpha _ke^{-j2\pi p\Delta f\tau _k}e^{j2\pi f_{D,k}qT_O}},
\\
&\forall (p,q) \in \varGamma _{s,k_A},
\end{aligned}
\label{Element_division}
\end{equation}
where $\boldsymbol{f}_{s,k_A}=\!\boldsymbol{f}_{s,q},q\in \varUpsilon _{k_A}$ is the transmit beamforming used for all the symbols in $\varUpsilon _{k_A}$.
Then, the periodogram is calculated to obtain the delays and Doppler shifts of the targets,
\begin{equation}
\begin{aligned}
\mathrm{Per}_{s,k_A}( \tau ,f_D ) &=\frac{1}{\left| \varGamma _{s,k_A} \right|}\sum_{\left( p,q \right) \in \varGamma _{s,k_A}}{\left[  \tilde{\boldsymbol{Y}}_{s,k_A} \right] _{p,q}}
\\
&\cdot e^{j2\pi p\Delta f\tau}e^{-j2\pi qT_Of_D}.
\end{aligned}
\label{Per_det_dedicated}
\end{equation}
The delay-Doppler pairs $\{(\hat{\tau}_k,\hat{f}_{D,k})\}_{k\in \Omega _{k_A}}$ of the targets can then be obtained by searching the spectrum $|\mathrm{Per}_{k_A}( \tau ,f_D )|^2$. The AoAs and delay-Dopplers are automatically paired, as the targets identified from $\mathrm{Per}_{k_A}( \tau ,f_D )$ has the same AoA estimate $(\hat{\phi}_{k_\phi},\hat{\theta}_{k_\phi})$. Then, since monostatic sensing is considered, the range and radial speed of the targets can be estimated as $\hat{r}_k=c\hat{\tau}_k/2$ and $\hat{v}_k=c\hat{f}_{D,k}/(2f_C)$ respectively, where $c$ denotes the speed of light and $f_C$ is the carrier frequency. Note that the beamforming design discussed in Section \ref{target_tracking} requires the knowledge of $\alpha_k$, which can be further estimated as
\begin{equation}
\begin{aligned}
\hat{\alpha}_k=\frac{\mathrm{Per}_{s,k_A}(\hat{\tau}_k,\hat{f}_{D,k})}{\boldsymbol{f}_{k_A}^{H}\hat{\boldsymbol{a}}_{k_A}\hat{\boldsymbol{a}}_{k_A}^T\!\boldsymbol{f}_{s,k_A}}.
\end{aligned}
\label{alpha_estimation}
\end{equation}

\subsubsection{Target Tracking with Dedicated Sensing Resources}
For target tracking, since no beam sweeping is conducted, and all the target angles are covered simultaneously by the tracking beamforming discussed in Section \ref{Tracking_dedicated}, the set of utilized TF resources is $\varGamma _s$, i.e., all the dedicated sensing resources are used to cover all the angles $\{\hat{\phi}_{k_A},\hat{\theta}_{k_A}\}_{k_A=0}^{K_A-1}$. The subsequent element-wise division and periodogram calculation are similar to \eqref{Element_division} and \eqref{Per_det_dedicated}, with the set $\varGamma _{s,k_A}$ replaced by $\varGamma _{s}$ and transmit beamforming $\boldsymbol{f}_{s,k_A}$ replaced by $\boldsymbol{f}_{s}$.

\subsubsection{Target Searching with Zero Sensing TF Resource}
For our proposed method, all the resources are used for sensing, but at angle $(\hat{\phi}_{k_A},\hat{\theta}_{k_A})$, only the corresponding symbols are utilized for delay-Doppler processing. Thus, the set of TF resources for target seatching at angle $(\hat{\phi}_{k_A},\hat{\theta}_{k_A})$ can be expressed as
\begin{equation}
\begin{aligned}
\varGamma _{k_A}\triangleq \left\{ \left( p,q \right) | q\in \varUpsilon _{k_A} \right\}.
\end{aligned}
\label{Resources_det_dedicated}
\end{equation}
Due to the use of different transmit beamforming for different users, the term $\boldsymbol{a}_k^T \boldsymbol{f}_{p,q}$ in \eqref{eq_mod_symb} is no longer a constant. Rather, it varies across the resources allocated to different users. Consequently, the equivalent modulation symbol $\beta _{k,{p},{q}}$ should be used instead of $b_{p.q}$ during element-wise division to completely remove the effects of modulation symbols. Since the value of $\beta _{k,{p},{q}}$ depends on the AoA $(\phi_k,\theta_k)$, an estimation of $\beta _{k,{p},{q}}$ for the targets in $\varOmega _{k_A}$ can be expressed as
\begin{equation}
\begin{aligned}
\hat{\beta}_{k_A,p,q}=\hat{\boldsymbol{a}}_{k_A}^T\boldsymbol{f}_{u,k_A}b_{p,q},
\end{aligned}
\label{Estimation_modulation_symbol}
\end{equation}
where $\boldsymbol{f}_{u,k_A} = \boldsymbol{f}_{u,q}, \, q \in \varUpsilon_{k_A}$ represents the transmit beamforming used for the symbols of the $u$th UE in the resource set $\varUpsilon_{k_A}$, and $u$ is the UE index corresponding to the resource element $(p,q)$. After performing element-wise division, the channel can be expressed as
\begin{equation}
\begin{aligned}
&\left[ \tilde{\boldsymbol{Y}}_{k_A} \right] _{p,q}=\frac{\left[ \boldsymbol{Y}_{k_A} \right] _{p,q}}{\hat{\beta}_{k_A,p,q}}
\\
&=\boldsymbol{f}_{k_A}^{H}\hat{\boldsymbol{a}}_{k_A}\sum_{k\in \varOmega _{k_A}}{\alpha _ke^{-j2\pi p\Delta f\tau _k}e^{j2\pi f_{D,k}qT_O}},\forall (p,q) \in \varGamma.
\end{aligned}
\label{Element_division_proposed}
\end{equation}
The periodogram is thus calculated as
\begin{equation}
\begin{aligned}
\mathrm{Per}_{k_A}(\tau ,f_D)=&\frac{1}{\left| \varGamma _{k_A} \right|}\sum_{\left( p,q \right) \in \varGamma _{k_A}}{\left[ \tilde{\boldsymbol{Y}}_{k_A} \right] _{p,q}}
\\
&\cdot e^{j2\pi p\Delta f\tau}e^{-j2\pi qT_Of_D},
\end{aligned}
\label{Per_det_proposed}
\end{equation}
and $\alpha_k$ can be estimated as
\begin{equation}
\begin{aligned}
\hat{\alpha}_k=\frac{\mathrm{Per}_{k_A}(\hat{\tau}_k,\hat{f}_{D,k})}{\boldsymbol{f}_{k_A}^{H}\hat{\boldsymbol{a}}_{k_A}}.
\end{aligned}
\label{alpha_estimation_proposed}
\end{equation}
\subsubsection{Target Tracking with Zero Sensing TF Resource}
In this case, all the resources are utilized for delay-Doppler processing. The element-wise division is conducted according to \eqref{Element_division_proposed}, with $\hat{\beta}_{k_A,p,q}$ estimated as
\begin{equation}
\begin{aligned}
\hat{\beta}_{k_A,p,q}=\hat{\boldsymbol{a}}_{k_A}^T\boldsymbol{f}_{u}b_{p,q},
\end{aligned}
\label{Estimation_modulation_symbol_2}
\end{equation}
where the transmit beamforming $\boldsymbol{f}_{u}$ is designed in Section \ref{Tracking_proposed}. The periodogram is then calculated according to \eqref{Per_det_proposed}, with $\varGamma _{k_A}$ replaced by $\varGamma$.

\section{Simulation Results} \label{sim_results}
In this section, we verify the effectiveness of the proposed beamforming design method and the spatial domain denoising method, and compare the communication and sensing performance between dedicated sensing resources and our proposed method. Unless otherwise specified, the simulation parameters are set as in Table \ref{tb:para}. The four users are assumed to have LoS channels with the BS, and are located at zenith and azimuth angles evenly distributed within $[30,150]^\circ$. The ranges of the users are uniformly distributed over $[10,100]\mathrm{m}$. 

\begin{table}[htb]
\caption{Parameter settings for simulation}
\centering
\label{tb:para}
\begin{tabular}{|c|c|c|c|c|}
\hline
\textbf{Symbols} & \textbf{Value}  \\
\hline
Carrier frequency $f_c$ & $28$ GHz\\
\hline
Subcarrier spacing $\Delta f$ & $120$ kHz\\
\hline
Number of subcarriers $P$ & 1024 \\
\hline
Bandwidth  & $122.88$ MHz \\
\hline
Number of OFDM symbols $Q$ & $1024$\\
\hline
CP length $T_{\mathrm{CP}}$ & $2.0833\ \mathrm{\mu s}$ \\
\hline
Noise power density $N_0$ & $-169$ dBm/Hz\\
\hline
Number of users $U$ & $4$\\
\hline
Number of subcarriers per user & 512\\
\hline
Number of OFDM symbols per user & 512\\
\hline
Number of sensing targets $K$ & 3\\
\hline
\end{tabular}
\end{table}
\subsection{Beamforming Design}
\begin{figure}[!htbp]
\vspace*{-6pt}
\centering
\includegraphics[width=0.4\textwidth]{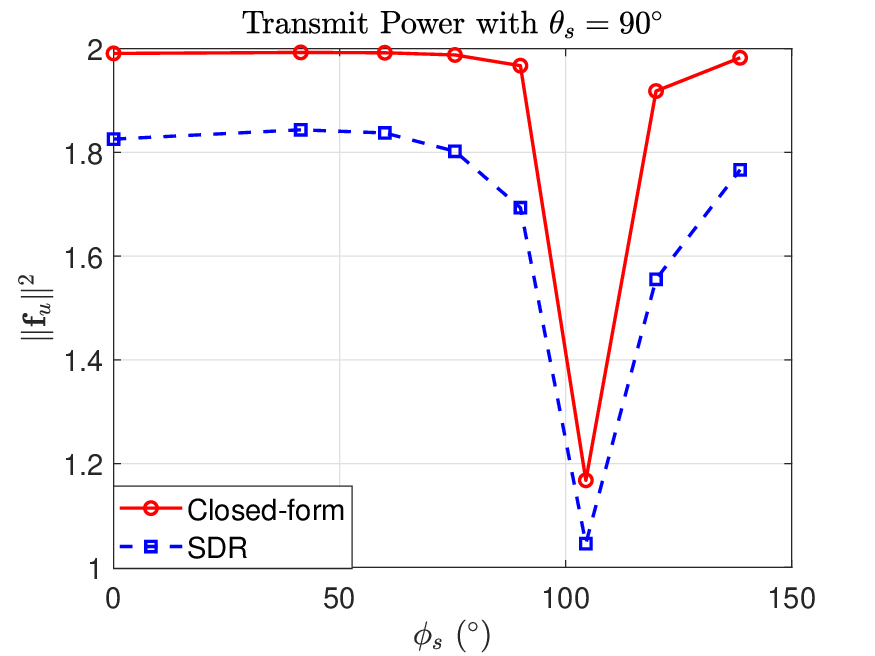}
\vspace*{-6pt}
\caption{Transmit power comparison between the optimal and closed-form beamforming.}
\label{pwr_comp}
\vspace*{-6pt}
\end{figure}

We first compare the performance of the low-complexity closed-form sub-optimal beamforming method presented in \eqref{Closed-Beamforming} with the optimal beamforming designed using SDR. In this comparison, we consider $ M \times N = 8 \times 8 $ transmit antennas. The user has a single LoS path with the BS at an AoD of $ (105,90)^\circ $, and we assume $ \mathcal{E}_{\min} = \bar{\gamma} \sigma^2 / |\lambda_u|^2 = 1 $ for simplicity.
In this configuration, the beamforming aims to transmit equal power toward both the sensing angle and the communication user. 
Figure \ref{pwr_comp} shows the transmit power for both beamforming schemes. The results indicate that both methods successfully achieve power sharing between sensing and communication. Moreover, the extra power transmitted by the closed-form beamforming is relatively minimal, demonstrating its effectiveness despite being sub-optimal.

\subsection{Denoising}
\begin{figure}[!htbp]
\centering
\subfloat[MUSIC spectrum without denoising]{
\includegraphics[width=0.5\linewidth]{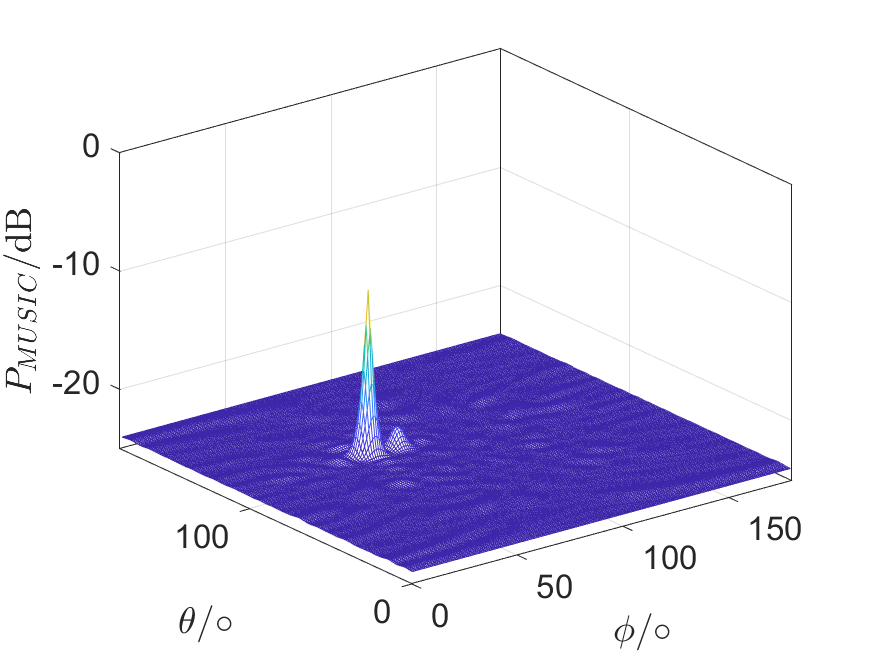}
\label{MUSIC_NoDenoise}
}
\subfloat[MUSIC spectrum with denoising]{
\includegraphics[width=0.5\linewidth]{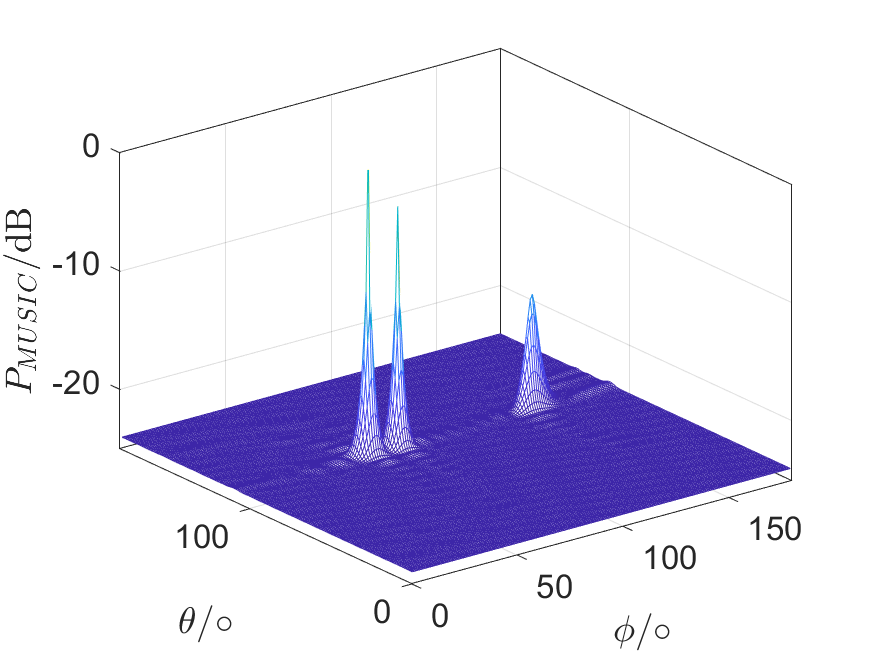}
\label{MUSIC_Denoise}
}
\caption{2D MUSIC spectrum (a) without and (b) with denoising.}
\label{DenoiseCompare}
\vspace*{-8pt}
\end{figure}

Next, we validate the performance of the spatial domain denoising. In the following simulation, $\bar{\gamma}$ is set to $0$ dB and $\mathcal{E} _{\min}$ is set to $-30$ dBm.
The sensing targets have AoAs $\{(63,109)^\circ,(143,112)^\circ,(77,109)^\circ\}$ respectively.
\begin{figure}[!htbp]
\centering
\subfloat[Euclidean norm histogram of \eqref{Denoise_Symbols}]{
\includegraphics[width=0.5\linewidth]{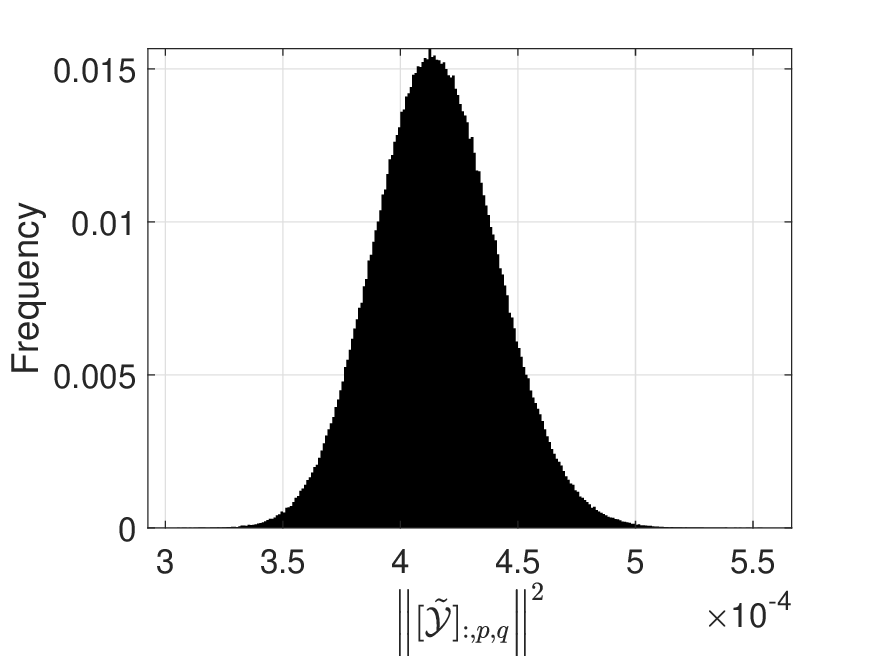}
\label{Hist_before}
}
\subfloat[Euclidean norm histogram of \eqref{DD_signal_searching}]{
\includegraphics[width=0.5\linewidth]{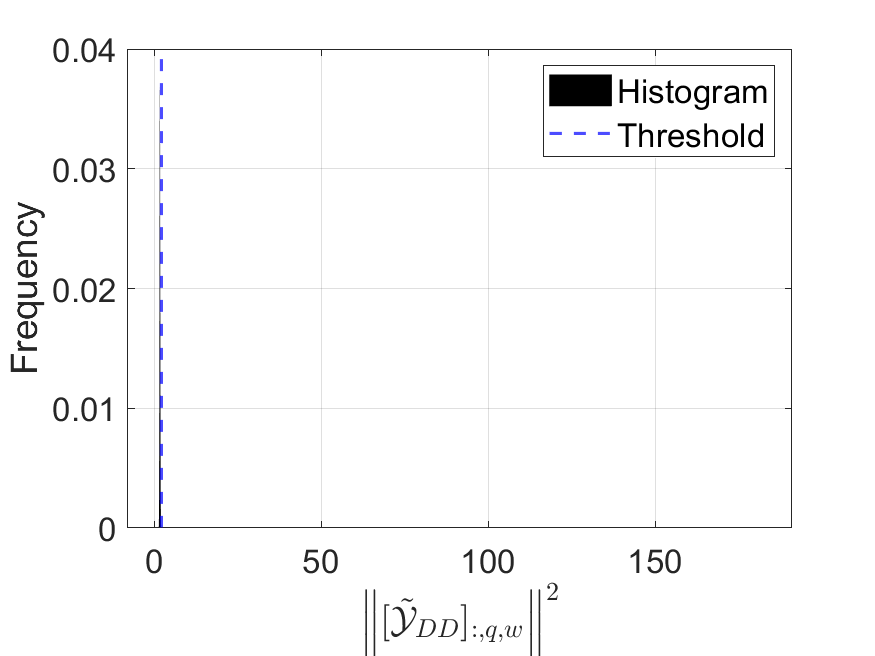}
\label{Hist_after}
}
\caption{Euclidean norm histogram of (a) $\tilde{\mathcal{Y}}$ in TF domain and (b) $\tilde{\mathcal{Y}}_{DD}$ in DD domain.}
\label{Hist}
\vspace*{-8pt}
\end{figure}
Fig. \ref{Hist} compares the Euclidean norm histogram of \eqref{Denoise_Symbols} and \eqref{DD_signal_searching}. As shown in the figure, in TF domain, the useful signal in $\tilde{\mathcal{Y}}$ is submerged in noise, and not directly separable. After transforming it to delay-Doppler domain, the useful signal becomes sparse and can be easily separated from the noise floor. The threshold $\mathcal{E} _{th}$ is chosen to retain the top $100$ points, as marked by the blue dashed line in Fig. \ref{Hist_after}. Fig. \ref{DenoiseCompare} compares the 2D MUSIC spectra with and without denoising. The original spectrum missed the target at $(143, 112)^\circ$ due to low SNR, while the denoised version successfully resolves all three target AoAs.

\subsection{Sensing Performance}\label{sensingperf}
In this subsection, we compare the sensing performance in the target sensing and tracking stages between the traditional approach with dedicated sensing resources and our proposed approach with zero TF sensing resource. The following simulations are conducted under two sets of target parameters shown in Table \ref{tb3}. In the far-apart scenario (F), all targets maintain sufficient spatial separation where both their AoA and delay differences are larger than the resolution limit of traditional and proposed systems. Conversely, in the close-proximity case (C), target $k=1$ and $k=3$ share the same AoA, and have narrow range separation so that the corresponding delay separation is smaller than the resolution limit of traditional systems with limited dedicated sensing resources but larger than that of the proposed approach.
\begin{table}[htbp]
	\centering  
	\caption{Parameters of close and far-apart sensing targets}  
	\label{tb3}  
	\begin{tabular}{c|c|c|c|c|c}  
		\hline  
		Case&$k$&Range $R_k$ (m)& RCS (dBsm)& $\phi_k$ ($^\circ$)&$\theta_k$ ($^\circ$) \\  
		\hline
		\multirow{3}{*}{Far apart (F)}&$1$&53.7&20&63&109 \\
            \cline{2-6}
      	~&$2$&125.7&20&143&112 \\
            \cline{2-6}
       	~&$3$&67.1&20&77&109 \\
		\hline
		\multirow{3}{*}{Close (C)}&$1$&53.7&20&63&109 \\
            \cline{2-6}
      	~&$2$&125.7&20&143&112 \\
            \cline{2-6}
       	~&$3$&56.2&20&63&109 \\
		\hline
	\end{tabular}
\end{table}

\subsubsection{Target Searching}
Here, we compare the sensing performance in the target searching stage between dedicated sensing resources and our proposed zero sensing resource approach. For a fair comparison, the transmit power of the two approaches are controlled the same by scaling $\bar{\gamma}$ and $\mathcal{E} _{\min}$. Note that for dedicated sensing resources, the power used for communication and sensing is expressed in \eqref{P_Tx,c} and \eqref{P_Tx,s} respectively. The communication-sensing power ratio for dedicated sensing resources can be expressed as,
\begin{equation}
\begin{aligned}
PR_D=\frac{(1-\kappa )\left\| \boldsymbol{f}_U \right\| ^2}{\kappa \left\| \boldsymbol{f}_S \right\| ^2}=\frac{\sigma ^2\bar{\gamma}_D}{\lambda _{u}^{2}\mathcal{E} _{\min ,D}}\cdot \frac{(1-\kappa )}{\kappa}.
\end{aligned}
\label{Pwr_ratio_dedicated}
\end{equation}
When the user channel $\boldsymbol{h}_u$ and the beam steering vector $\boldsymbol{a}_{s,q}$ are orthogonal, the communication-sensing power ratio for our proposed method is
\begin{equation}
\begin{aligned}
PR_P=\frac{\sigma ^2\bar{\gamma}_P}{\lambda _{u}^{2}\mathcal{E} _{\min ,P}}.
\end{aligned}
\label{Pwr_ratio_proposed}
\end{equation}
As a result, in order to keep the communication-sensing power ratios the same, the ratios of $\bar{\gamma}$ and $\mathcal{E} _{\min}$ should be adjusted so that
\begin{equation}
\begin{aligned}
\frac{\bar{\gamma}_P}{\mathcal{E} _{\min ,P}}=\frac{(1-\kappa )\bar{\gamma}_D}{\kappa \mathcal{E} _{\min ,D}}.
\end{aligned}
\label{thres_ratio_adjust}
\end{equation}
In the following simulations, $\mathcal{E} _{\min,P}/\bar{\gamma}_P$ is kept at $-30~\mathrm{dBm}$, with $\mathcal{E} _{\min,D}/\bar{\gamma}_D$ adjusted according to \eqref{thres_ratio_adjust} and the transmit beamforming for our proposed method is designed with SDR.

\begin{figure}[!htbp]
\centering
\subfloat[Number of detected targets]{
\includegraphics[width=0.4\textwidth]{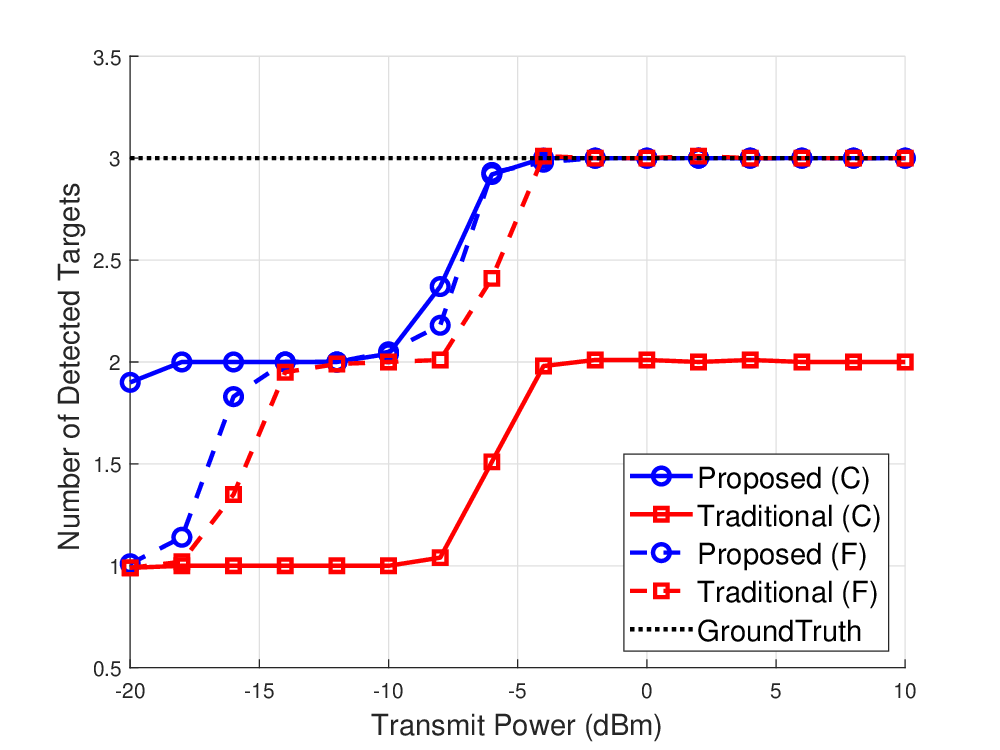}
\label{SearchNum}
}

\subfloat[RMSE of azimuth angle estimations]{
\includegraphics[width=0.5\linewidth]{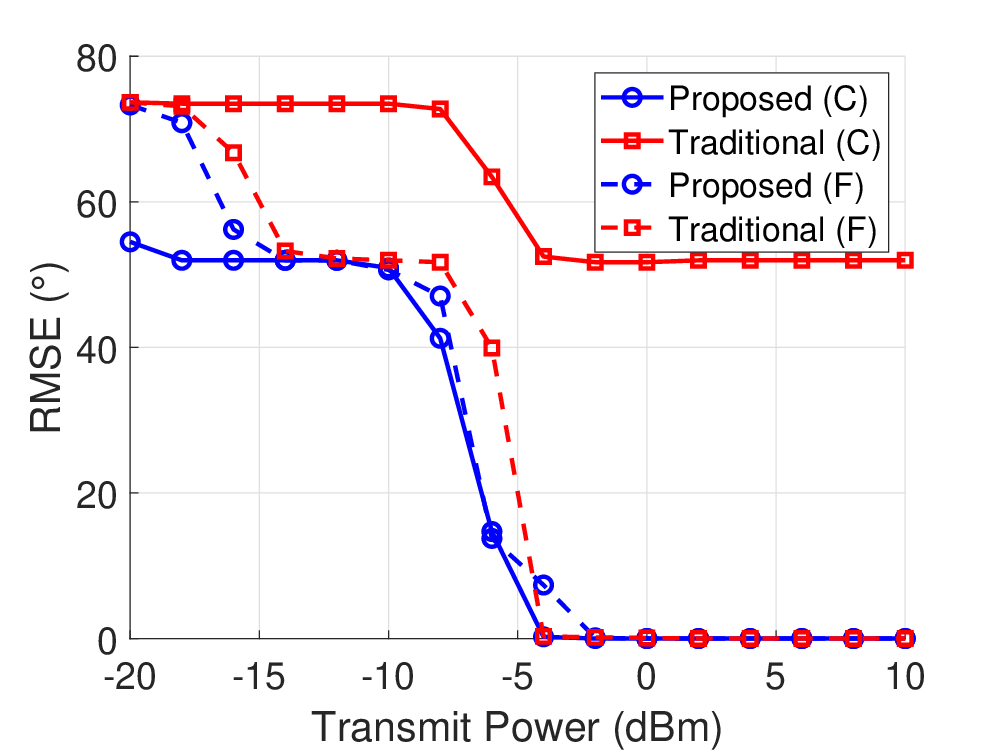}
\label{SearchAngle}
}
\subfloat[RMSE of delay estimations]{
\includegraphics[width=0.5\linewidth]{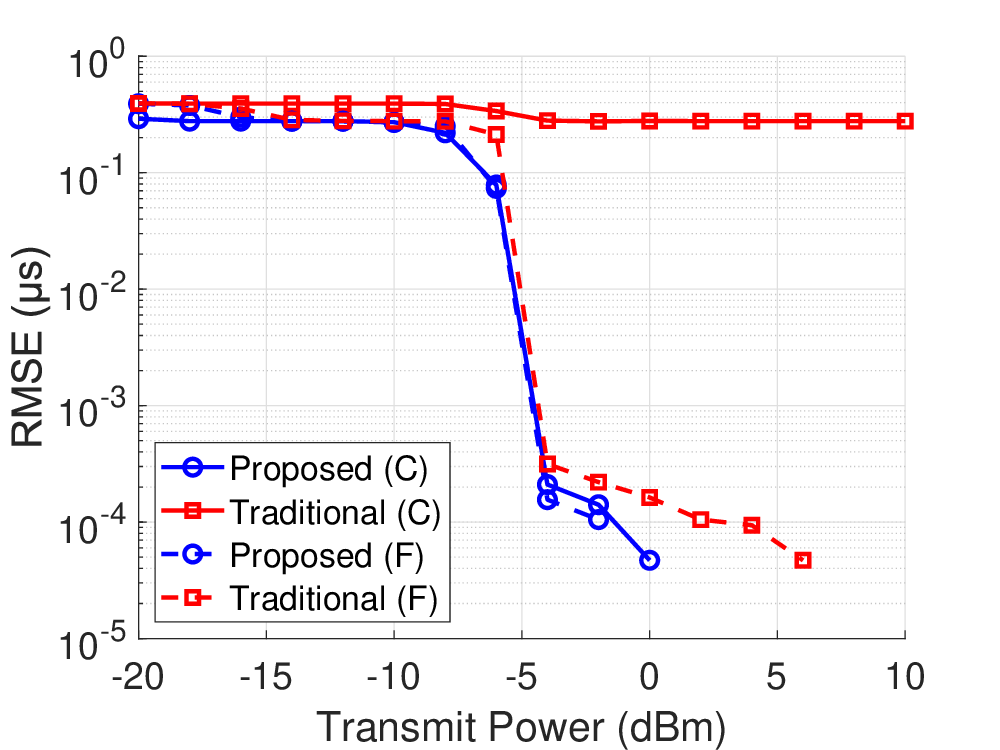}
\label{SearchDelay}
}
\caption{Comparison of (a) number of detected targets (b) RMSE of azimuth angle estimations and (c) RMSE of delay estimations between in target searching stage traditional approach with dedicated sensing resources and our proposed approach with zero TF sensing resource.}
\label{SearchingCompare}
\vspace*{-8pt}
\end{figure}

Fig. \ref{SearchingCompare} compares the sensing performance of the traditional and proposed approach under the close and far-apart scenarios.
As shown in Fig. \ref{SearchNum}, in the far-apart scenario, the traditional and proposed approach have similar detection performance. For the close-proximity case, the traditional approach fails to resolve the closely distributed targets due to limited sensing resources while the proposed approach successfully detects the three targets. Fig. \ref{SearchAngle} shows the root mean square error (RMSE) of azimuth angle estimations versus the transmit power. The results for zenith angles are similar and are omitted for brevity. Note that the angle estimation errors for undetected targets are set to $90^\circ$. As shown in the figure, in the far-apart scenario, the angle sensing performance is similar for both approaches while in the close-proximity scenario, the proposed method outperforms the traditional method due to success in resolving all three targets.

For delay estimations, the proposed method outperforms the traditional approach in two aspects. As illustrated in Fig. \ref{SearchDelay}, for the far-apart case, the proposed method demonstrated superior accuracy. This is attributable to the lower CRLB provided by increased TF resources. For the close-proximity case, the performance gap further amplifies, where the traditional method fails to detect one of the targets. Note that similar to angle estimation, the delay estimation errors for undetected targets are set to the delay spread among all sensing targets.

\subsubsection{Target Tracking}

Next, we compare the sensing performance in the target tracking stage. In the following simulations, the transmit power of the two approaches are controlled the same. Similar to target sensing, the communication and sensing SNR constraint ratios $\bar{\gamma}/\bar{\gamma}_s$ are adjusted to ensure same communication-sensing power ratio for both approaches,
\begin{equation}
\begin{aligned}
\frac{\bar{\gamma}_P}{\bar{\gamma}_{s,P}}=\frac{(1-\kappa )\bar{\gamma}_D}{\kappa \bar{\gamma}_{s,S}}.
\end{aligned}
\label{thres_ratio_adjust_track}
\end{equation}
In the following simulations, ${\bar{\gamma}_P}/{\bar{\gamma}_{s,P}}$ is kept at $30~\mathrm{dB}$.

\begin{figure}[!htbp]
\centering
\subfloat[Number of detected targets]{
\includegraphics[width=0.4\textwidth]{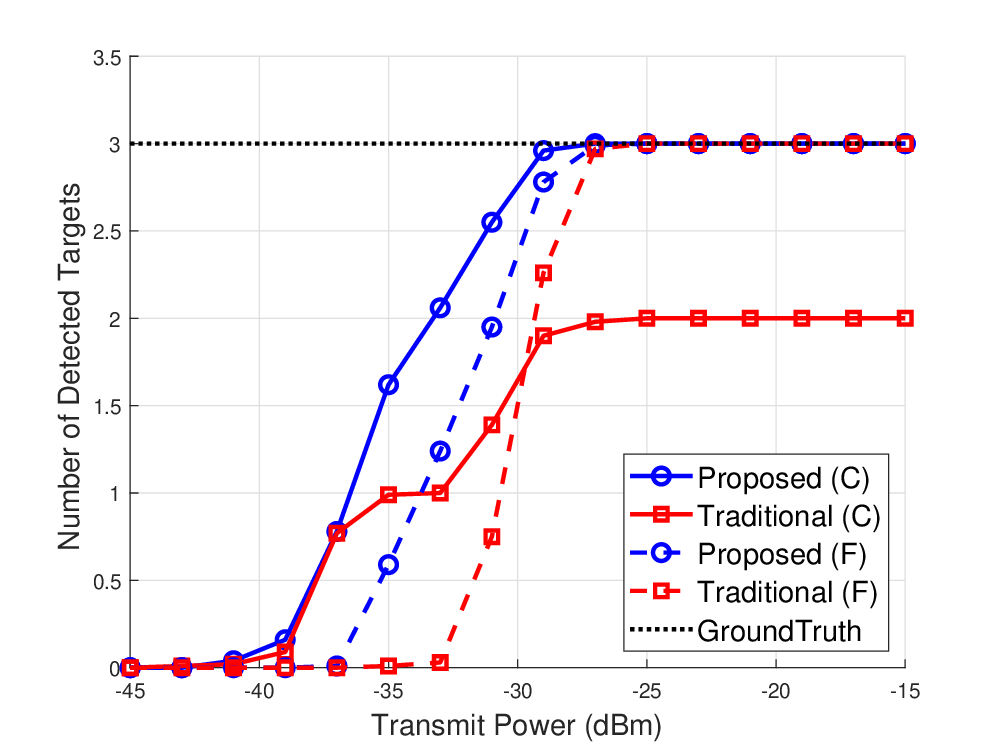}
\label{TrackNum}
}

\subfloat[RMSE of azimuth angle estimations]{
\includegraphics[width=0.5\linewidth]{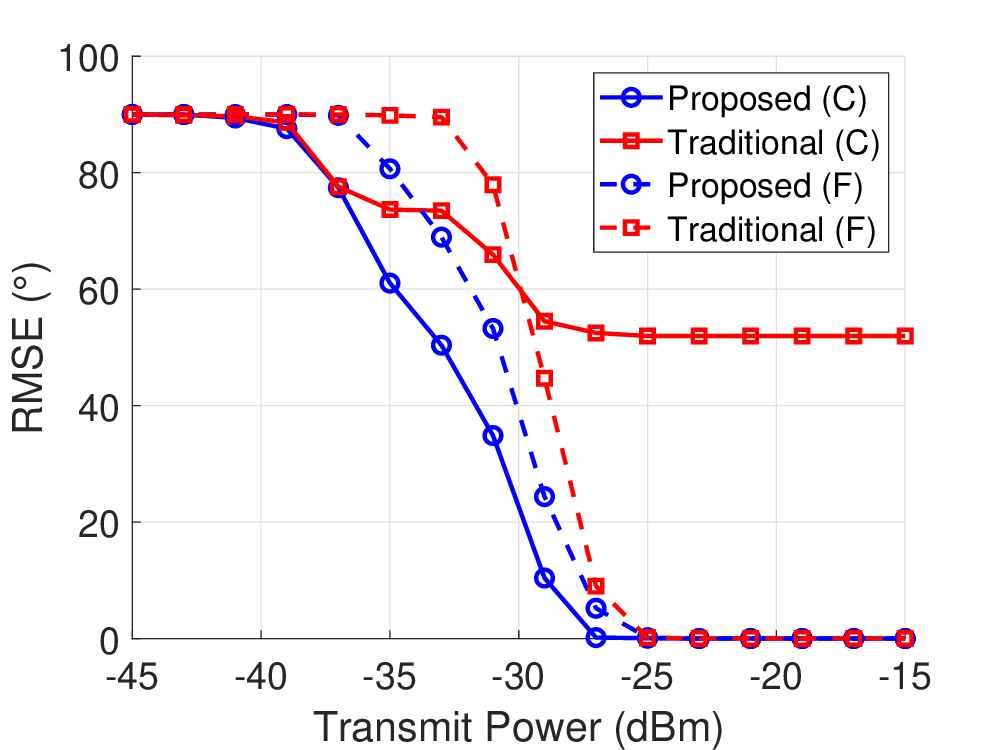}
\label{TrackAngle}
}
\subfloat[RMSE of delay estimations]{
\includegraphics[width=0.5\linewidth]{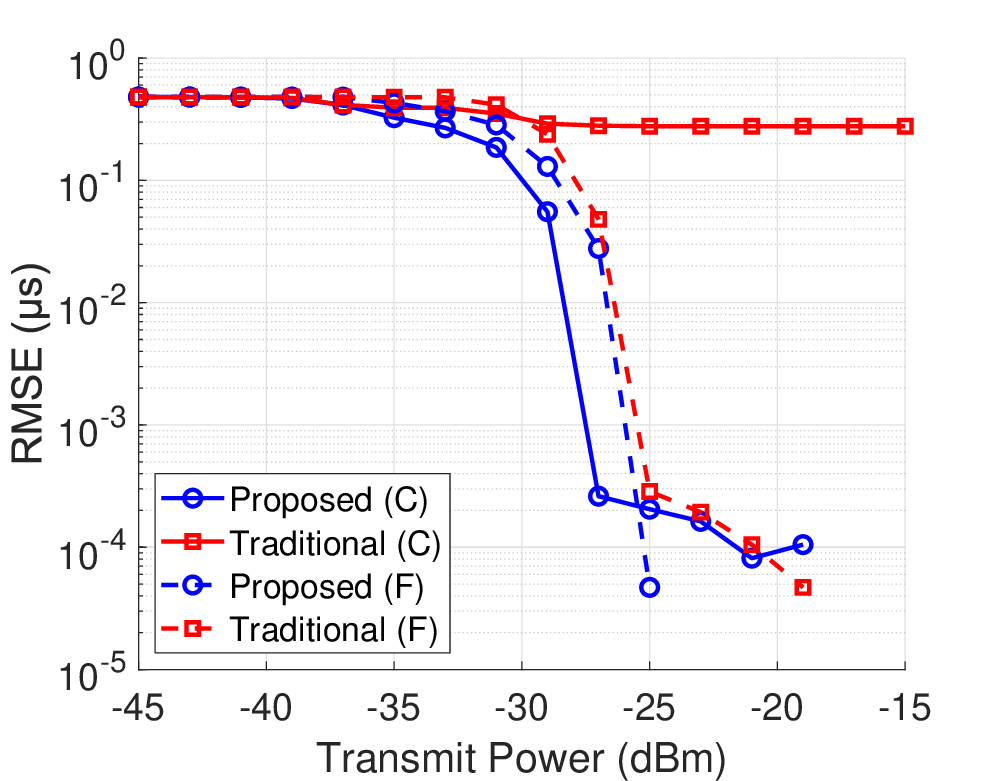}
\label{TrackDelay}
}
\caption{Comparison of (a) number of detected targets (b) RMSE of azimuth angle estimations and (c) RMSE of delay estimations in target tracking stage between traditional approach with dedicated sensing resources and our proposed approach with zero TF sensing resource.}
\label{TrackingCompare}
\vspace*{-8pt}
\end{figure}

As illustrated in Fig. \ref{TrackingCompare}, in the target tracking stage, similar to target searching, the detection performance of two approaches is similar in the far-apart scenario while the proposed method demonstrates superior resolution in the close-proximity case. For angle sensing, the two approaches also has similar performance in the far-apart scenario while the proposed method achieves higher accuracy in the close-proximity case thanks to better resolution. For delay sensing, similar to target searching, the proposed method achieves higher accuracy in both scenarios thanks to the lower CRLB and better resolution.

By comparing Fig. \ref{SearchingCompare} and Fig. \ref{TrackingCompare}, we can observe that, to obtain similar sensing performance, target tracking requires significantly (around $25~\mathrm{dB}$) less  power than target searching, thanks to the transmit beamforming towards the targets of interest. This demonstrates superior energy efficiency of our proposed two-stage sensing procedure.

\subsection{Communication Performance}
In this subsection, we compare the communication performance between the traditional and proposed approaches. The following simulations are conducted under parameters shown in Table \ref{tb4}.

\begin{table}[htbp]
	\centering  
	\caption{Parameters of Communication Users}  
	\label{tb4}  
	\begin{tabular}{c|c|c|c|c|c}  
		\hline  
		$u$ & $\tau_u$ (ns) & $f_{D,u}$ (kHz) & $|\alpha_u|$ & $\phi_u$ ($^\circ$)&$\theta_u$ ($^\circ$)\\  
		\hline
		$1$&113.9&2.5&$2.4\times 10^{-5}$&106&41 \\
            \hline
      	$2$&325.5&3.2&$8.8\times 10^{-6}$&96&145 \\
            \hline
       	$3$&317.4&-3.0&$8.9\times 10^{-6}$&49&147 \\
		\hline
      	$4$&73.2&3.3&$3.7\times 10^{-5}$&88&120 \\
		\hline
	\end{tabular}
\end{table}

Note that user $u=4$ is used only for the proposed approach, and is allocated the TF resources otherwise dedicated for sensing with the traditional approach.

\begin{figure}[!htbp]
\centering
\subfloat[Sum rates in the searching stage.]{
\includegraphics[width=0.5\linewidth]{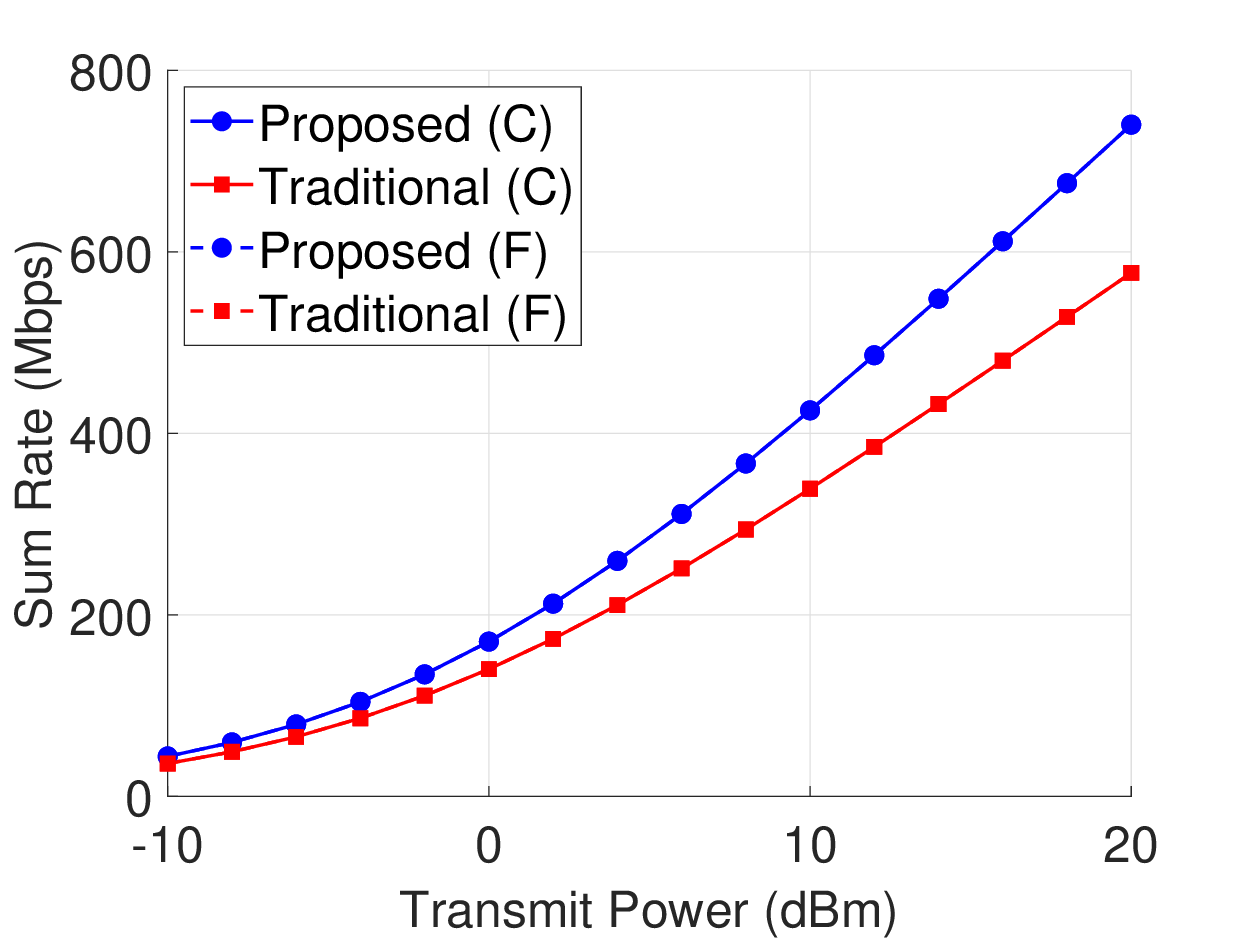}
\label{com_search}
}
\subfloat[Sum rates in the tracking stage.]{
\includegraphics[width=0.5\linewidth]{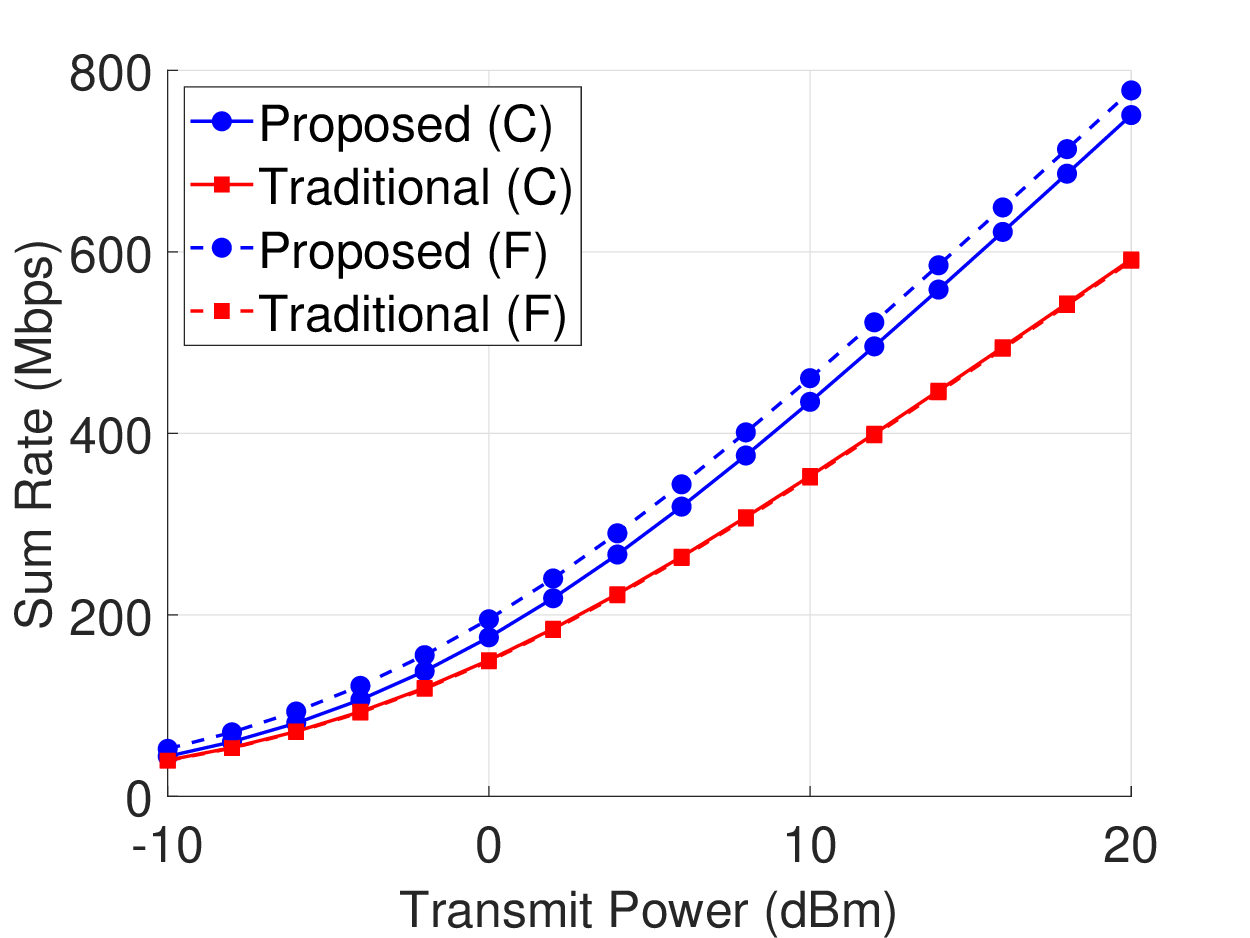}
\label{com_track}
}
\caption{Comparison of  communication rates in the (a) searching stage and (b) tracking stage between traditional approach with dedicated sensing resources and our proposed approach with zero TF sensing resource.}
\label{CommunicationCompare}
\vspace*{-8pt}
\end{figure}


Fig. \ref{com_search} compares the communication performance in the searching stage. Thanks to the additional TF resources available for communication, our proposed method achieves significantly higher sum rates than the traditional method. Note that the sum rates overlap in the close-proximity and far-apart cases described in Table \ref{tb3}, which is expected since the beamforming does not depend on target locations in the searching stage.
Fig. \ref{com_track} compares the communication performance in the tracking stage. Similar to the searching stage, our proposed method achieves significantly higher sum rates than the traditional method. Note that the sum rates overlap in the close-proximity and far-apart cases for the traditional approach. This is because the communication-sensing power ratio is fixed, and thus not affected by target distribution. For our proposed approach, however, when the targets are closely located, the sum rate is higher, this is because less additional power is required to satisfy the sensing SNR requirements and more power is thus available for communication.

\section{Conclusion} \label{conclusion}
In this paper, we proposed a novel multi-user MIMO OFDM ISAC framework for low-altitude UAVs with zero sensing resource allocation. Different from the traditional approach with dedicated sensing TF resources, our proposed approach reuses the communication TF resources for sensing by designing the transmit beamforming to meet both communication and sensing requirements. The beamforming design strategies and sensing algorithms for target searching and tracking stages were developed. Simulation results validated the superior communication and sensing performance of our proposed approach in terms of sum rate, sensing accuracy and resolution.

\bibliographystyle{IEEEtran}
\bibliography{IEEEabrv,reference}

\end{document}